\documentclass[11pt]{article}

\usepackage{amsmath,amsthm,amsfonts,amssymb,bm}
\usepackage{natbib}
\usepackage{graphicx,subcaption}
\usepackage{enumerate}
\usepackage[dvipsnames]{xcolor}
\usepackage{tikz}
\usepackage{booktabs,multirow}
\usepackage[margin=1in]{geometry}
\usepackage{microtype}
\usepackage{url}
\usepackage{setspace}
\def\bSig\mathbf{\Sigma}

\def\diag{{\mathrm{diag}}}

\def\Ox{\Omega_X}
\def\Oy{\Omega_Y}
\def\HS{_{\mathrm{HS}}}

\def\c{\mathcal}

\def\simiid{\overset{i.i.d.}{\sim}}
\def\dcd{\,\cdot\,,\,}

\def\hSigzd{\hat\Sigma_{\tilde Z\tilde Z}^{\dagger}}
\def\hSigdzx{\hat\Sigma_{\tilde Z (\tilde X \tilde Z)}}

\def\hSigdyx{\hat\Sigma_{\tilde Y (\tilde X \tilde Z)}}
\def\hSigdyz{\hat\Sigma_{\tilde Y\tilde Z}}
\def\hSigz{\hat{\Sigma}_{\tilde Z\tilde Z}}
\def\hSigtx{\hat\Sigma_{\tilde X\tilde X}}

\def\hSigty{\hat\Sigma_{\tilde Y\tilde Y}}

\def\Sigxyz{\Sigma_{ \tilde Y   (\tilde X \tilde Z) | \tilde Z}}
\def\hSigxyz{\hat\Sigma_{ \tilde Y   (\tilde X \tilde Z) | \tilde Z}}

\def\sigt{\Sigma_{\tilde Y\tilde X}}
\def\hsigt{\hat\Sigma_{\tilde Y\tilde X}}
\def\Gx{G_{(\tilde X\tilde Z)\mid\tilde Z}}
\def\Gy{G_{\tilde Y\mid\tilde Z}}
\def\epsn{\epsilon_n}

\def\ka{\kappa}

\DeclareMathOperator{\supp}{supp}
\def\cD{\overset{\c D}{\rightarrow}}
\def\cP{\overset{P}{\rightarrow}}
\def\Gyx{\Gamma_{\tilde Y\tilde X}}
\def\Gyxz{\Gamma_{\tilde Y(\tilde X\tilde Z)\mid\tilde Z}}
\def\Sigz{\Sigma_{\tilde Z\tilde Z}}
\def\lx{\lambda_X}
\def\ly{\lambda_Y}
\def\lz{\lambda_Z}
\def\bigtimes{\mbox{\LARGE{$\times$}}}

\def\real{\mathbb{R}}
\def\ran{\mathrm{ran}}
\def\cran{\overline{\ran}}
\def\ker{\mathrm{ker}}
\def\E{\mathbb{E}}
\def\R{\mathbb R}

\def\cov{\mathrm{Cov}}
\def\tr{\mathrm{tr}}
\def\eD{\overset{\c D}{=}}

\theoremstyle{plain}
\newtheorem{theorem}{Theorem}

\newtheorem{corollary}[theorem]{Corollary}
\newtheorem{assumption}{Assumption}
\newtheorem{definition}{Definition}

\newcommand{\indep}{\;\, \rule[0em]{.03em}{.6em} \hspace{-.25em}
\rule[0em]{.65em}{.03em} \hspace{-.25em}
\rule[0em]{.03em}{.6em}\;\,}

\begin{document}

\title{\vspace{-1.5cm}Distance Profile Embedding for Independence and Conditional Independence Testing of Random Objects}
\author{
Wenxi Tan, Bing Li, and Lingzhou Xue\\[0.3em]
Department of Statistics, The Pennsylvania State University\\
\texttt{wkt5100@psu.edu}, \texttt{bxl9@psu.edu}, \texttt{lzxue@psu.edu}
}
\date{}
\maketitle

\begin{abstract}
Testing independence or conditional independence is fundamental to statistical inference, yet existing methods for non-Euclidean random objects often face a difficult trade-off between geometric flexibility and theoretical tractability. We introduce the Distance Profile Embedding (DPE), a novel representation that maps random objects from general metric spaces into a Hilbert space of square-integrable functions. We prove that this mapping is injective and preserves full distributional information without requiring isometric Hilbert embeddings or one-to-one correspondence conditions. Leveraging the DPE, we develop a unified framework for marginal and conditional independence testing of random objects that enjoys a rigorous asymptotic theory for both size and power. Notably, our framework is the first in the literature to accommodate object-valued conditioning variables when testing conditional independence, overcoming the Euclidean or Hilbertian constraints of existing methodologies. We facilitate the calculation of analytic $p$-values using closed-form asymptotic null distributions, which avoids the computational burden of permutation tests common in existing metric-based methods. The numerical properties of our methods are demonstrated through both simulations and two real-world applications involving gut microbiome compositions and global human mortality distributions, respectively.
\end{abstract}


\section{Introduction}
Testing independence and conditional independence is essential to statistical inference, forming the basis for dimension reduction, variable selection, and causal discovery.
Traditional nonparametric testing methods have been well developed for Euclidean data \citep{Hoeffding1948, szekely2007measuring, sejdinovic2013equivalence, shao2014martingale, wang2015conditional, zhu2020distance,fan2024test}. These approaches rely fundamentally on linear structure, natural ordering, or the existence of characteristic functions that are intrinsic to Euclidean spaces.

Modern scientific studies, however, increasingly involve complex non-Euclidean data (also known as ``random objects") that take values in general metric spaces, with representative examples including phylogenetic trees, microbiome compositions, brain networks, and probability measures in Wasserstein space.
The need for rigorous inference about independence and conditional independence for such objects arises naturally across a wide range of scientific domains.
In microbial ecology, for example, testing the association between the microbiome composition and the outcomes of interest is a primary problem \citep{COMBO2011Gary}. The compositional nature of these data and their underlying phylogenetic relationships often require the use of tree-based metrics, such as UniFrac \citep{li_microbiome_2015}, which capture biological similarity more effectively than the standard Euclidean distance \citep{zhao2015testing}. The inherent difficulty of this problem lies in the lack of a vector space structure, rendering the analytical machinery of classical inference inapplicable. Because methods based on ranks, projections, or characteristic functions cannot be directly extended to random objects, new theoretical frameworks are required to accommodate the intrinsic geometry of general metric spaces.

A prevailing strategy for analyzing random objects is to embed the metric space isometrically into a Hilbert space, thereby enabling the use of classical linear and kernel-based tools. In the context of independence testing, the validity of distance covariance (dCov) requires the metric space to be of strong negative type, which necessarily admits an isometric Hilbert embedding \citep{szekely2007measuring}. Similarly, the Hilbert-Schmidt Independence Criterion (hSIC) consistently detects dependence only when the underlying kernel is characteristic \citep{gretton2007kernel}. Standard kernel choices, such as distance-induced or Gaussian-type kernels, are characteristic primarily under strong negative type assumptions \citep{sejdinovic2013equivalence, ziegel2024characteristic}.  This reliance on strong negative type limits the scope of these methods. Notably, many important non-Euclidean domains, including spheres equipped with geodesic distance, multi-dimensional Wasserstein spaces,
and phylogenetic trees \citep{bhattacharjee2025doubly},
are not of strong negative type and do not admit Hilbert space embeddings.

To circumvent the restrictive requirement of isometric embedding into Hilbert spaces, a recent line of work represents the distribution of a random object through its {\em distance profile}---the collection of distances from the object to a set of reference points in the metric space. As shown in Table \ref{tab:marginal_method_comparison}, notable examples include the Ball Covariance (Ball) \citep{wang_nonparametric_2024}, the Distance Profile-based Mutual Independence Test (DiPMInd) \citep{chen2024testing}, and the Profile Association (PA) \citep{zhou2025association}. These methods broaden the applicability of independence testing in metric spaces by reducing inference on complex objects to real-valued distance variables. However, their theoretical guarantees rely on a stringent structural assumption of \emph{one-to-one correspondence} \citep{Christensen1970, Hoffmann1975}, which requires that the collection of distance summaries uniquely determines the underlying probability measure (see Section \ref{sec: DP} for details).   This condition is often difficult to verify and can fail in general metric spaces, as demonstrated by a counterexample in \cite{Davies_1971}. The need for an alternative framework that preserves full distributional information without requiring such identifiability conditions leads to an important open problem:

\begin{center}
\emph{($Q1$) Could we construct an independence test for random objects in general metric spaces without requiring isometric Hilbert embeddings or one-to-one correspondence conditions?}
\end{center}
\begin{table}[ht]
    \centering
    \renewcommand{\arraystretch}{0.9}
    \caption{Marginal independence testing methods for random objects in metric spaces. \textit{Assumption} denotes the structural requirement for consistency; \textit{Asymptotics} reports asymptotic size control and consistency for the marginal test; \textit{Inference} indicates analytic or permutation-based $p$-value calculation.}
    \label{tab:marginal_method_comparison}
    \resizebox{\textwidth}{!}{
    \begin{tabular}{lcccc}
        \toprule
        \multirow{2}{*}{\textbf{Method}} & \multirow{2}{*}{\textbf{Assumption}} & \multicolumn{2}{c}{\textbf{Asymptotics}} & \multirow{2}{*}{\textbf{Inference}} \\
        \cmidrule(lr){3-4}
        & & \textbf{Size} & \textbf{Consistency} & \\
        \midrule
        hSIC \citep{gretton2007kernel} & Hilbert & $\checkmark$ & $\times$ & Permutation \\
        dCov \citep{szekely2007measuring} & Hilbert & $\checkmark$ & $\times$ & Permutation \\
        DiPMInd \citep{chen2024testing} & one-to-one corresp. & $\checkmark$ & $\checkmark$ & Permutation \\
        Ball \citep{wang_nonparametric_2024} & one-to-one corresp. & $\checkmark$ & $\times$ & Permutation / Analytic \\
        PA \citep{zhou2025association} & one-to-one corresp. & $\checkmark$ & $\checkmark$ & Permutation \\
        \textbf{DPE (Ours)} & None above & $\checkmark$ & $\checkmark$ & Analytic \\
        \bottomrule
    \end{tabular}
    }
\end{table}

The structural and theoretical constraints of existing frameworks become even more pronounced in the context of conditional independence, a concept central to many statistical areas such as graphical modeling and causal discovery. While the current literature on random objects focuses on characterizing marginal associations, learning dependence structures in graphical models or isolating direct causal effects requires conditioning on variables that may themselves be random objects. This challenge is exemplified in global health, particularly in investigating the relationship between childbearing and female survival \citep{WHO2019MaternalMortality}. Rigorous evaluation of the dependence between fertility and mortality distributions requires controlling for shared socioeconomic and environmental confounders. A principled strategy involves conditioning on the male mortality distribution, which serves as a proxy for these latent factors. However, the development of systematic procedures for independence tests conditioning on such infinite-dimensional random objects remains an important open problem.

\begin{table}[ht]
    \centering
    \renewcommand{\arraystretch}{0.9}
    \caption{Conditional independence testing methods for random objects in metric spaces. \textit{Target Assumption} and \textit{Cond. Assumption} denote the structural requirements on the target and conditioning variables, respectively; \textit{Asymptotics} reports asymptotic size control and consistency for the conditional test; \textit{Inference} indicates analytic or permutation-based $p$-value calculation.}
    \label{tab:conditional_method_comparison}
    \resizebox{\textwidth}{!}{
    \begin{tabular}{lccccc}
        \toprule
        \multirow{2}{*}{\textbf{Method}} & \multirow{2}{*}{\textbf{Target Assumption}} & \multirow{2}{*}{\textbf{Cond. Assumption}} & \multicolumn{2}{c}{\textbf{Asymptotics}} & \multirow{2}{*}{\textbf{Inference}} \\
        \cmidrule(lr){4-5}
        & & & \textbf{Size} & \textbf{Consistency} & \\
        \midrule
        KCI \citep{zhang2012kernel} & Hilbert & Hilbert & $\checkmark$ & $\times$ & Analytic \\
        PA \citep{zhou2025association} & one-to-one corresp. & Euclidean & $\checkmark$ & $\checkmark$ & Permutation \\
        \textbf{DPE (Ours)} & None above & None above & $\checkmark$ & $\checkmark$ & Analytic \\
        \bottomrule
    \end{tabular}
    }
\end{table}

Existing frameworks for conditional independence testing impose stringent structural conditions that limit their generalizability. For instance, the Profile Association (PA) test \citep{zhou2025association} restricts the conditioning variable to Euclidean
spaces. {The Kernel Conditional Independence (KCI) test \citep{zhang2012kernel} relies on characteristic kernels that, as previously noted, require Hilbert structures for all variables. Adding to the structural constraints, KCI lacks universal consistency } in the alternative for conditional tests involving general metric spaces, as summarized in Table \ref{tab:conditional_method_comparison}. The pervasive need to overcome these geometric and theoretical bottlenecks leads to a subsequent fundamental question:

\begin{center}
\noindent\emph{($Q2$) Could we construct a conditional independence test for random objects without requiring isometric Hilbert embeddings or Euclidean structures?}
\end{center}

In this paper, we address both ($Q1$) and ($Q2$) by developing a unified framework for testing the independence or conditional independence of random objects in general metric spaces. We introduce a novel representation, termed the \textbf{Distance Profile Embedding (DPE)}, which maps each random object to an integrable function on the metric space defined by its distances to reference points. This representation induces a Hilbertian structure at the level of the embedded objects, facilitating the assessment of dependence via cross-covariance operators in the reproducing kernel Hilbert spaces (RKHS). Crucially, unlike existing approaches, the DPE preserves the full distributional information of the original random objects without requiring isometric Hilbert embeddings or one-to-one correspondence conditions, thereby resolving ($Q1$). Furthermore, it addresses ($Q2$) by accommodating object-valued conditioning variables within an RKHS-based conditional independence framework.

To translate this DPE representation into a practical inferential tool, we establish the large-sample theory and computational framework necessary for hypothesis testing. We develop computationally efficient statistics with analytically tractable asymptotic distributions under the null hypothesis of independence or conditional independence, avoiding the need for computationally intensive permutation procedures. The theoretical consistency and power of our methods are established through rigorous proofs and validated via numerical simulations on metric spaces that do not admit isometric Hilbert embeddings. Finally, we demonstrate the practical utility of our framework through applications to complex real-world datasets, including gut microbiome compositions and human mortality distributions.

More specifically, we summarize our contributions as follows:
\begin{itemize}
    \item \textbf{Distance Profile Embedding for General Metric Spaces.} The DPE is a novel mapping from a general metric space to a Hilbert space of square-integrable functions. We prove that the DPE is injective and measurable, thus preserving the full distributional information of the original random objects without requiring the underlying metric space to admit an isometric Hilbert embedding (see Theorem \ref{theorem:joint measurability} and Corollaries \ref{corollary:independence}--\ref{corollary:conditional independence}). Beyond the scope of independence and conditional independence testing, the DPE provides a principled bridge between random objects in general metric spaces and classical Hilbert-space theory.

    \item \textbf{Independence Testing for Random Objects without One-to-One Correspondence.}
      We address the open problem ($Q1$) by leveraging the DPE representation to develop a general framework for independence testing that is free from restrictive structural conditions. Recent methods \citep{wang_nonparametric_2024, zhou2025association} rely heavily on the one-to-one correspondence assumption \citep{Christensen1970, Hoffmann1975}, which may not hold as pointed out by \cite{wang_nonparametric_2024}.

      \item \textbf{Conditional Independence Testing with Object-Valued Conditioning Variables.}
    To the best of our knowledge, this is the first framework for conditional independence testing that operates when all variables, including the conditioning variable, are random objects in metric spaces. Unlike existing approaches that require Euclidean covariates \citep{zhou2025association} or Hilbertian structures for all variables \citep{zhang2012kernel}, our method accommodates object-valued conditioning without requiring such restrictive geometries, resolving the fundamental question ($Q2$).

    \item \textbf{Comprehensive Asymptotic Theory for Size and Power.}
    As shown in Tables \ref{tab:marginal_method_comparison} and \ref{tab:conditional_method_comparison}, while asymptotic size control and power consistency have been established in some settings (e.g., \cite{chen2024testing} and \cite{zhou2025association}),
    a unified theory, particularly for testing conditional independence, has remained unaddressed. We provide the asymptotic theory that includes: (i) central limit theorems for our proposed measures of independence and conditional independence; (ii) explicit asymptotic characterizations of the test statistics under both null and alternative hypotheses; and (iii) theoretically justified $p$-value calculations and local power analyses.
    \item  \textbf{Analytic Inference without Resampling.} Existing nonparametric tests for random objects, such as \cite{chen2024testing, zhou2025association}, often rely on permutation-based procedures, which are computationally prohibitive in high-dimensional or large-scale settings. Our framework enables analytic inference by providing test statistics with tractable asymptotic null distributions. This achieves significant computational savings while maintaining numerical stability, precise size control, and competitive power across various regimes.
\end{itemize}

The remainder of the paper is organized as follows. Section~\ref{sec: DP} reviews distance profiles and discusses their limitations. Section~\ref{sec: Th DPE} formalizes our DPE framework and establishes its fundamental distributional properties. Section~\ref{sec: UI} develops our DPE-based independence testing procedures, providing the associated asymptotic theory and implementation details. Section~\ref{sec: CI} extends the framework to conditional independence testing and presents a rigorous asymptotic analysis for the object-valued setting. Section~\ref{sec: referen measure} discusses the selection and construction of reference measures in practice. Section~\ref{sec: numerics} evaluates the finite-sample performance of our methods in simulation studies, and Section~\ref{sec: rda} demonstrates the practical utility through applications to gut microbiome compositions and human mortality distributions. Section~\ref{conclusion} includes a few concluding remarks. All technical proofs are presented in the supplement.

\section{Preliminaries} \label{sec: DP}

\def\lo{_}
\def\ca#1{{\cal #1}}
\def\of{\circ}
\def\inv{^{-1}}

We introduce some basic notation and assumptions used in this paper.

\smallskip

\textbf{Random objects.} Let $(\Omega, \ca F, P)$ be a probability space and let $\Ox$ and $\Oy$ be two complete separable metric spaces, equipped with metrics $d_X(\cdot, \cdot)$ and $d_Y(\cdot, \cdot)$, respectively. Let $\ca F \lo X$ and $\ca F \lo Y$ be the Borel $\sigma$-fields generated by open sets in $\Omega \lo X$ and $\Omega \lo Y$, respectively. Let $X: \Omega \to \Omega \lo X$ be a random element measurable with respect to $\ca F / \ca F \lo X$, and $Y: \Omega \to \Omega \lo Y$ measurable with respect to $\ca F/ \ca F \lo Y$. Then, $(X, Y): \Omega \to \Omega \lo X \times \Omega \lo Y$ is measurable with respect to $\ca F / ( \ca F \lo X \times \ca F \lo Y)$, where $\ca F \lo X \times \ca F \lo Y$ is the product $\sigma$-field. Let $P \lo X = P \of X \inv$, $P \lo Y = P \circ Y \inv$, and $P \lo {XY} = P \of (X, Y)\inv$ be the distributions of $X$, $Y$, and $(X,Y)$, respectively.

For each $u \in \Omega \lo X$ and  $v \in \Omega \lo Y$, we assume that $d \lo X (u, X) $ and $d \lo Y (v, Y)$ have finite first moments:
$\int \lo {\Omega \lo X} d_X(u, x) dP_X(x) < \infty$, and $\int  \lo {\Omega \lo Y} d_Y(v, y) dP_Y(y) < \infty$.
We use $X \eD X^\prime$ to indicate that random objects $X$ and $X^\prime$ are identically distributed. The support of a measure $\lambda$ on a metric space $\c M$ is defined as the set
$\supp (\lambda)=\{x\in \c M: \lambda(\c I(x))>0\text{ for every neighborhood $\c I(x)$ of $x$}\}$.
For a random element $R$ on $(\Omega, \ca F, P)$, we also use $\supp (R)$ to denote the support of the distribution of $R$; that is, $\supp (R) = \supp (P \of R \inv)$.

\smallskip

\textbf{Distance Profile.} The distance profile introduced by \cite{dubey2024metrics} will be important for our development. Formally, the distance profile of a random object $X$ at a fixed location $u\in \Omega_X$ is defined as the cumulative distribution function (CDF) of the distance $d_X(u, X)$, that is,
$
F^X_u(t) = P(d_X(u, X)\le t), \; t \ge 0.
$
The collection $\{F^X_u: u \in \Omega_X\}$ forms a family of univariate distribution functions. Distance profiles have emerged as a pivotal tool for object-valued inference \citep{chen2024testing, zhou2025association}, providing an interpretable summary of the underlying distribution of object-valued data by leveraging the marginals of the stochastic process $\{ d_X(u, X): u \in \Omega_X \}$.

\smallskip

\textbf{The One-to-One Correspondence Condition.} A critical structural assumption in current profile-based inference is that the collection of probabilities assigned to all closed metric balls, specifically, $P(d_X(X, u)\le r)$ for all $u\in \Ox\,, r \ge 0$, uniquely characterizes the underlying probability measure $P_X$. This condition implies that two random objects $X$ and $X'$ taking values in $\Omega_X$ are identically distributed if and only if $d_X(u,X)$ and $d_X(u,X')$ are identically distributed for all $u \in \Omega_X$. This property, called the \emph{one-to-one correspondence}, has its roots in early foundational work on the geometry of measures \citep{Christensen1970, Hoffmann1975}.

However, this correspondence does not hold in general. \cite{Davies_1971} provided a counterexample by constructing distinct probability measures that assign identical probabilities to all closed metric balls, thereby inducing indistinguishable distance profiles. Despite this fundamental limitation, most existing methods for object-valued independence testing rely on assumptions that presume this one-to-one correspondence; see Section S1.1 of the Supplementary Material for more details. While the one-to-one correspondence provides a convenient bridge between object-valued independence and the factorization of distance profiles, its lack of universality and the inherent difficulty of its verification motivate the development of a new framework that offers rigorous theoretical guarantees without imposing such restrictive geometric conditions.

{\section{Distance Profile Embedding}\label{sec: Th DPE}
This section introduces the distance profile embedding (DPE) to encode a random object as a function in a Hilbert space. Unlike the previously developed embedding-based methods, our embedding does not impose additional structure on the metric space, such as a negative-type metric or a Riemannian manifold. We begin with DPE for a single random object, and then move to DPE of several random objects. The multiple-object embedding is needed for the subsequent development because the independence test involves two random objects, and the conditional independence test involves three random objects. The main theoretical issue involved in this construction is measurability: we need the embedded objects to be measurable with respect to appropriate product $\sigma$-fields so that independence and conditional independence can be built on a firm theoretical foundation.

\subsection{Distance Profile Embedding of a Single Random Object}

\def\hi{^}

Let $\lambda \lo X$ be a measure on $(\Omega \lo X, \ca F \lo X)$, which we call {\em the reference measure}.
It can be, but need not be, the probability measure.  Let $L \lo 2 (\Omega \lo X, \lambda \lo X) $ be the class of  real-valued measurable functions on $\Omega \lo X$ that are square-integrable with respect to $\lambda \lo X$, which is a Hilbert space with an inner product
$\langle f_1, f_2 \rangle_{\lambda_X} = \int f_1(u) f_2(u) \, d\lambda_X(u)$;
that is,
$L \lo 2 (\Omega_X, \lambda \lo X)
= \{ f : \  \int \lo {\Omega \lo X} f(u)^2 \, d\lambda_X(u) < \infty \}$.
\begin{assumption}\label{assumption:d in L2}
    For each $x \in \Omega \lo X$, $d \lo X (\cdot, x) \in L \lo 2 (\Omega \lo X, \lambda \lo X) $.
\end{assumption}

We make the above assumption. Under Assumption \ref{assumption:d in L2}, we define the map $\Phi \lo X: \Omega \lo X \to L \lo 2 (\Omega \lo X, \lambda \lo X) $  by
    $\Phi \lo X (x) = d \lo X ( \cdot, x)$.
That is, for each $x \in \Omega \lo X$, $\Phi \lo X(x)$ is the function $u \mapsto d \lo X (u, x)$. We use $[\Phi \lo X(x)](u)$, or simply $\Phi \lo X (x)(u)$, to denote the real number $d \lo X (u, x)$. Letting $X: \Omega \to \Omega \lo X$ be a random object in $(\Omega \lo X, \ca F \lo X)$, we formally define the DPE of $X$.
\begin{definition} The distance profile embedding (DPE) of $X$ is the mapping
\begin{align*}
   \Phi \lo X (X): \Omega \to L \lo 2 (\Omega \lo X, \lambda \lo X) , \quad \omega \mapsto \Phi \lo X (X (\omega)).
\end{align*}
\end{definition}
As we will show in Subsection \ref{subsection:measurability}, in a more general setting, $\Phi \lo X(X)$ is a random element in $L \lo 2 (\Omega \lo X, \lambda \lo X)$ under mild conditions.

\subsection{Distance Profile Embedding of Several Random Objects}
Let $(\Omega \lo {X \hi i}, d \lo {X \hi i}): i = 1, \ldots, m$ be $m$ metric spaces. For example, in Section \ref{sec: UI}, $m=2$, and in Section \ref{sec: CI}, $m=3$. We use superscripts to label the metric spaces and reserve the subscript for indexing the subjects later.   For $i=1, \ldots, m$, let $\ca F \lo {X \hi i}$ be a $\sigma$-field on the set $\Omega \lo {X \hi i}$, and $\lambda \lo {X \hi i}$ be a measure on $\ca F \lo {X \hi i}$.

\def\ali{&\, \, }

\begin{assumption}\label{assumption:d in L2 several}
   For each $i=1, \ldots, m$ and  $x \hi i \in \Omega \lo {X \hi i}$, $d \lo {X \hi i}(\cdot, x \hi i) \in L \lo 2 ( \Omega \lo {X \hi i}, \lambda \lo {X \hi i})$.
\end{assumption}
We make the above assumption.   Under Assumption \ref{assumption:d in L2 several}, we define the map
\begin{align*}
\Phi \lo {X \hi 1 \cdots X \hi m}:   \Omega\lo {X \hi 1} \times \cdots \times \Omega \lo {X \hi m} \ali \to L \lo 2 ( \Omega \lo {X \hi 1}, \lambda \lo {X \hi 1} ) \times \cdots \times L \lo 2 ( \Omega \lo {X \hi m}, \lambda \lo {X \hi m} ), \\
 (x \hi 1, \ldots, x \hi m) \ali \mapsto ( d \lo {X \hi 1} (\cdot, x \hi 1), \ldots, d \lo {X \hi m} ( \cdot, x \hi m)).
\end{align*}
Let
    $(X \hi 1, \ldots, X \hi m) : \Omega \to \Omega \lo {X \hi 1} \times \cdots \times \Omega \lo {X \hi m}$
be a multivariate random object measurable with respect to the product $\sigma$-field $\ca F \lo {X \hi 1} \times \cdots \times \ca F \lo {X \hi m}$.
 We now define the DPE of $(X \hi 1, \ldots, X \hi m)$.
\begin{definition} The distance profile embedding (DPE) of $(X \hi 1, \ldots, X \hi m)$ is the mapping
\begin{align*}
   \Phi \lo {X \hi 1 \cdots X \hi m} (X \hi 1, \ldots, X \hi m): \Omega \ali \to L \lo 2 (\Omega \lo {X \hi 1}, \lambda \lo {X \hi 1 }) \times \cdots \times L \lo 2 ( \Omega \lo {X \hi m} , \lambda \lo {X \hi m}), \\
   \omega \ali \mapsto (\Phi \lo {X \hi 1}(X \hi 1 (\omega)), \ldots, \Phi \lo {X \hi m} (X \hi m (\omega))).
\end{align*}
\end{definition}

Note that this definition does not guarantee that  $\Phi \lo {X \hi 1 \cdots X \hi m} (X \hi 1, \ldots, X \hi m)$ is a random object in the product measurable space, but this is true under mild conditions, as we will show in the next subsection.

\subsection{Injectivity, Continuity and Measurability}\label{subsection:measurability}
For each $i=1, \ldots, m$, let $\ca T \lo {X \hi i}$ be the topology on $\Omega \lo {X \hi i}$ induced by the metric $d \lo {X \hi i}$, $\ca F \lo {X \hi i}$ a $\sigma$-field on $\Omega \lo {X \hi i}$, and $\lambda \lo {X \hi i}$ a measure on $(\Omega \lo {X \hi i}, \ca F \lo {X \hi i})$. Let $\ca T \lo {\tilde X \hi i}$ be the topology on $L \lo 2 (\Omega \lo {X \hi i}, \lambda \lo {X \hi i})$ induced by its inner product, and $\ca F \lo {\tilde X \hi i}$ a $\sigma$-field on $L \lo 2 (\Omega \lo {X \hi i}, \lambda \lo {X \hi i})$. For $k$ generic topological spaces $\ca T \lo 1, \ldots, \ca T \lo k$, let $\ca T \lo 1 \times \cdots \times \ca T \lo k$ denote their product topology. For $k$ generic $\sigma$-fields $\ca F \lo 1, \ldots, \ca F \lo k$, let $\ca F \lo 1 \times \cdots \times \ca F \lo {k}$ denote their product measurable space. For $k$ generic measures $\mu \lo 1, \ldots, \mu \lo k$, let $\mu \lo 1 \times \cdots \times \mu \lo {k}$ denote their product measure. For a topology $\ca T$, let $\ca B (\ca T)$ be the Borel $\sigma$-field generated by $\ca T$.

Let $A$ be a subset of $\{1, \ldots, m \}$. Let
 $\Omega \lo {X \hi A}   = \underset{i \in A}{\bigtimes} \Omega \lo {X \hi i}$,  $\ca T \lo {X \hi A}   = \underset{i \in A}{\bigtimes} \ca T \lo {X \hi i}$,  $\ca F \lo {X \hi A}   = \underset{i \in A}{\bigtimes} \ca F \lo {X \hi i}$, and $\lambda  \lo {X \hi A}   = \underset{i \in A}{\bigtimes} \lambda \lo {X \hi i}$. Similarly, let
$\ca T \lo {\tilde X \hi A}   = \underset{i \in A}{\bigtimes} \ca T \lo { \tilde X \hi i}$ and  $\ca F \lo {\tilde X \hi A}   = \underset{i \in A}{\bigtimes} \ca F \lo {\tilde X \hi i}$ be the product topology and product $\sigma$-field on the Cartesian product
 $\underset{i \in A}{\bigtimes} L \lo 2 (\Omega \lo {X \hi i}, \lambda \lo {X \hi i})$. In the following, we use $(v \lo i: i \in A)$ to denote the vector corresponding to the set $\{v \lo i: i \in A\}$. There is a subtle difference between the vector $(v \lo i : i \in A )$ and the set $\{v \lo i: i \in A \}$. For example, if $A = \{1, 2, 5\} $. Then $(v \lo i: i \in A )$ is the vector $(v \lo 1, v \lo 2, v \lo 5)$, which is an ordered sequence of members of the set $\{v \lo 1, v \lo 2, v \lo 5  \}$. Mathematically, a vector is a mapping from $i \in A$ to $\{v \lo i: i \in A \}$, which is slightly different from the set $\{v \lo i: i \in A \}$ itself: a set is the range of the map that defines the corresponding vector.
Using this vector notation, we define $\Phi \lo {X \hi A}$ as the mapping
\begin{align*}
    \Phi \lo {X \hi A}: \ali  \Omega \lo {X \hi A} \to \underset{i \in A} {\bigtimes} L \lo 2 ( \Omega \lo {X \hi i}, \lambda \lo {X \hi i}), \quad
      x \hi A \mapsto (\Phi \lo {X \hi i} (x \hi i): i \in A ).
\end{align*}
We introduce a metric in $\bigtimes \lo {i \in A} L \lo 2 ( \Omega \lo {X \hi i}, \lambda \lo {X \hi i})$ as
\begin{align*}
    d \lo {\tilde X \hi A} (\tilde x \lo 1 \hi A, \tilde x \lo 2 \hi A) = \sum \lo {i \in A}   \| \tilde x \lo 1 \hi i - \tilde x \lo 2 \hi i \| \lo {L \lo 2 (\Omega \lo {X \hi i}, \lambda \lo {X \hi i})}.
\end{align*}
It is easy to verify that this metric generates the product topology $\ca T \lo {\tilde X \hi A}$. We also introduce $d_{X^A}$ on $\bigtimes \lo {i \in A}  \Omega \lo {X \hi i}$ analogously.
There are many topologically equivalent metrics on $\bigtimes \lo {i \in A} L \lo 2 (\Omega \lo {X \hi i}, \lambda \lo {X \hi i})$, but the above is convenient and sufficient for our discussion.
In the following, let $\mathrm{supp}(\mu)$ denote the support of a measure $\mu$. The next theorem lays out some fundamental properties of the mapping $\Phi \lo {X \hi A}$.

\begin{theorem}\label{theorem:joint measurability} Suppose Assumption \ref{assumption:d in L2 several} is satisfied and let $A$ be a subset of $\{1, \ldots, m \}$.
\begin{enumerate}
\item If \ $\supp ( \lambda \lo {X \hi i}) = \Omega \lo {X \hi i}$ \  for each $i=1, \ldots, m$, then the mapping $\Phi \lo {X \hi A}$ is injective in terms of the metrics $d \lo {X \hi A}$ and $d \lo {\tilde X \hi A}$.
    \item If, for each $i = 1, \ldots, m$,  $\lambda \lo {X \hi i}$ is a finite measure, then  $\Phi \lo {X \hi A}$  is continuous with respect to the product topologies  $\ca T \lo {X \hi A}$ and $\ca T \lo {\tilde X \hi A}$.
    \item If, in addition, $\Omega \lo {X \hi i}$ is a Polish metric space and $\ca F \lo {X \hi i} = \ca B ( \ca T \lo {X \hi i})$, $\ca F \lo {\tilde X \hi i} = \ca B ( \ca T \lo {\tilde X \hi i})$ for $i =1, \ldots, m$, then $\Phi \lo {X \hi A}$ is measurable with respect to the product $\sigma$-fields $\ca F \lo {X \hi A}$ and $\ca F \lo {\tilde X \hi A}$.
\end{enumerate}
\end{theorem}

\def\hi{^}

Thus, under the conditions in Theorem \ref{theorem:joint measurability}, $\Phi \lo {X  \hi A } (X \hi A)$ is a random element taking values in $\bigtimes \lo {i \in A} \,  L \lo 2 (\Omega \lo {X \hi i}, \lambda \lo {X \hi i}) $. A realization of this random element is $\Phi \lo {X \hi A } ( X \hi A  (\omega))$, a member of $\bigtimes \lo {i \in A} \, L \lo 2 (\Omega \lo {X \hi i}, \lambda \lo {X \hi i}) $. The evaluation of this realization at $x \hi A$ is $\Phi \lo {X \hi A } (X \hi A (\omega)) (x \hi A )$, which is a real (and nonnegative at all elements) vector.
Two important special cases of $\Phi \lo {X \hi A}$ are obtained when $A = \{i\}$ is a singleton and when $A=\{1,\ldots,m\}$; these correspond to the DPE of a single random object $X \hi i$ and the joint DPE of $(X \hi 1, \ldots, X \hi m)$, respectively.
Note that, via DPE, we can turn any random object (or vector of random objects) into a Hilbert-space valued random function (or a vector of Hilbert-space valued functions) without requiring the metric space $\Omega \lo X$ to be of negative type. The cost of not having the negative-type condition is that we do not have an isometric embedding, but measurability and injectivity are all we need to preserve independence or conditional independence as we move from metric spaces to Hilbert spaces. Below, we summarize the conditions made in Theorem \ref{theorem:joint measurability} as the next assumption.

\begin{assumption}\label{assumption:for dpe} For each $i = 1, \ldots, m$, $\Omega \lo {X \hi i}$ is a Polish metric space, $\lambda \lo {X \hi i}$ is a finite measure such that $\supp(\lambda_{X^i}) = \Omega \lo {X \hi i}$, $\ca F \lo {X \hi i} = \ca B ( \ca T \lo {X \hi i})$  and $\ca F \lo {\tilde X \hi i} = \ca B ( \ca T \lo {\tilde X \hi i})$.
\end{assumption}

It is also worth noting that the distance profile introduced by \citet{dubey2024metrics} relies on the total boundedness of $\Ox$ as a prerequisite, which is a topological constraint consistently imposed in recent literature \citep{chen2024testing, wang_nonparametric_2024, zhou2025association}. With mild conditions in Assumptions \ref{assumption:d in L2 several}--\ref{assumption:for dpe}, our DPE framework extends the applicability of profile-based inference to metric spaces that may not be totally bounded.}

\subsection{Preservation of Independence and Conditional Independence}\label{subsection:preserving}

In this subsection, we show that DPE preserves independence and conditional independence, so that we can equivalently test them in the original metric spaces or the embedded Hilbert spaces. Intuitively, these equivalences hold because independence and conditional independence are inherent properties of $\sigma$-fields, and measurable and injective mappings do not change   $\sigma$-fields.

Let $X$ and $Y$ be random objects defined on $(\Omega \lo X, d \lo X)$ and $(\Omega \lo Y, d \lo Y)$, and $(\tilde X, \tilde Y)$ the DPE of $(X, Y)$. From Theorem \ref{theorem:joint measurability} we know that, under Assumptions \ref{assumption:d in L2 several} and \ref{assumption:for dpe}, $(\tilde X, \tilde Y)$ is a random element taking values in the measurable space $(L \lo 2 ( \Omega \lo X, \lambda \lo X) \times L \lo 2 (\Omega \lo Y, \lambda \lo Y), \ca F \lo {\tilde X} \times \ca F \lo {\tilde Y})$.

\begin{corollary}\label{corollary:independence} Under Assumptions \ref{assumption:d in L2 several}--\ref{assumption:for dpe}, if $(X, Y)$ is a random object taking values in the product measurable space $(\Omega \lo X \times \Omega \lo Y, \ca F \lo X \times \ca F \lo Y)$ and $(\tilde X, \tilde Y)$ is DPE of $(X, Y)$, then $X \indep Y$ if and only if $\tilde X \indep \tilde Y$.
\end{corollary}

Now let $Z$ be a third random object defined on a metric space $(\Omega \lo Z, d \lo Z)$. Under Assumptions \ref{assumption:d in L2 several} and \ref{assumption:for dpe}, $(\tilde X, \tilde Y, \tilde Z)$ is a random object taking values in the product measurable space  $\left( L \lo 2 (\Omega \lo X, \lambda \lo X) \times  L \lo 2 (\Omega \lo Y, \lambda \lo Y) \times L \lo 2 (\Omega \lo Z, \lambda \lo Z), \ \ca F \lo {\tilde X} \times \ca F \lo {\tilde Y} \times \ca F \lo {\tilde Z} \right)$.

\begin{corollary}\label{corollary:conditional independence} Under Assumptions \ref{assumption:d in L2 several}--\ref{assumption:for dpe}, if $(X, Y, Z)$ is a random object taking values in the product measurable space
   $ ( \Omega \lo X \times \Omega \lo Y \times \Omega \lo Z, \ \ca F \lo X \times \ca F \lo Y \times \ca F \lo Z)$ and $(\tilde X, \tilde Y, \tilde Z)$ is the DPE of $(X, Y, Z)$, then $X \indep Y |Z $ if and only if $\tilde X \indep \tilde Y | \tilde Z$.
\end{corollary}

\section{Construction of DPE-Based Independence Test} \label{sec: UI}
Built on Corollary~\ref{corollary:independence}, we now develop a principled and computationally tractable testing procedure for the independence between the Hilbertian random elements $\tilde X$ and $\tilde Y$.

\def\hi{^}

\subsection{DPE-Based Characterization of Independence}

Let $\ka_{\tilde X}$ and $\ka_{\tilde Y}$ be continuous kernels on $L \lo 2 (\Omega \lo X, \lambda \lo X)$ and $L \lo 2 (\Omega \lo Y, \lambda \lo Y)$, and let $\c H_{\tilde X}$ and $\ca H \lo {\tilde Y}$, respectively,  be the reproducing kernel Hilbert spaces (RKHS) generated by them. Following \cite{gretton2007kernel} and \cite{li2018linear}, let  $\sigt: \c H_{\tilde X} \rightarrow \c H_{\tilde Y}$ be the following operator
\begin{equation}\label{eq:Sigma XY}
    \sigt= \E[\ka_{\tilde Y}(\dcd \tilde Y) \otimes \ka_{\tilde X}(\dcd \tilde X) ] - \E[\ka_{\tilde Y}(\dcd \tilde Y)] \otimes \E[\ka_{\tilde X}(\dcd \tilde X)],
\end{equation}
where, for example, for $\tilde x \in L \lo 2(\Ox, \lx)$, $\ka \lo {\tilde X} (\cdot, \tilde x)$ is the function $\tilde u \mapsto \ka \lo {\tilde X }(\tilde u, \tilde x)$, and for $f \in \ca H \lo {\tilde X}$ and $g \in \ca H \lo {\tilde Y}$, $g \otimes f$ is the linear operator defined by
   $ \ca H \lo {\tilde X} \to \ca H \lo {\tilde Y}, \ h \mapsto g \langle f, h \rangle \lo {\ca H \lo {\tilde X}}$.
Furthermore, the expectation in (\ref{eq:Sigma XY}) is defined by Bochner's integral (see \cite{hsing2015theoretical}).  Under the assumption
    $\E  [ \ka \lo {\tilde X} (\tilde X,  \tilde X) \hi {1/2} \, \ka \lo {\tilde Y} (\tilde Y,  \tilde Y) \hi {1/2}  ] < \infty$,
$\Sigma \lo {\tilde Y \tilde X}$ is a bounded operator satisfying
$
\langle g, \sigt f\rangle_{\c H_{\tilde Y}}  = \cov [ f(\tilde X), g (\tilde Y )]
$
for any $f \in \ca H \lo {\tilde X}$ and $g \in \ca H \lo {\tilde Y}$. For this reason, $\Sigma \lo {\tilde Y \tilde X}$ is called the cross-covariance operator. We assume a slightly stronger condition
as follows.
\begin{assumption}\label{assumption:kernel finite}
   $\E[\ka_{\tilde X}(\tilde X, \tilde X)] < \infty$, and $\E[\ka_{\tilde Y}(\tilde Y, \tilde Y)] < \infty$.
\end{assumption}

A kernel $\ka$ is said to be {\em characteristic} if it uniquely determines a distribution. More specifically, if $\ka$ is a positive definite kernel and $\ca H$ is the RKHS generated by it, then  $\ka$ is characteristic if, for two random elements $X$ and $X'$, $\E f(X) = \E f(X')$ for all $f \in \ca H$ implies $X \overset{\ca D}{=} X'$ (see \cite{zhang2012kernel}).
When $\ka_{\tilde X}$ and $\ka_{\tilde Y}$ are characteristic, $\sigt$ is the zero operator if and only if $\tilde X \indep \tilde Y$, providing a criterion for independence testing in the embedded space \citep{gretton2007kernel}. We record this fact in the next theorem.

\begin{theorem}\label{thm: equiv of idp}
Suppose that Assumptions \ref{assumption:d in L2 several}, \ref{assumption:for dpe}, and \ref{assumption:kernel finite}   are satisfied. If $\ka_{\tilde X}$ and $\ka_{\tilde Y}$ are  characteristic, then  $\tilde X \indep \tilde Y$ if and only if
 $\sigt=0$.
\end{theorem}

By Theorem 3.1 of \cite{ziegel2024characteristic}, the Gaussian kernel on a separable Hilbert space is a canonical example of a characteristic kernel (see also \cite{zhang2024dimension,zhang2024nonlinear} for more examples). This yields the following corollary, which is important for the implementation of our proposed framework.

\begin{corollary}\label{cor: RBF UI} Suppose Assumptions \ref{assumption:d in L2 several} and  \ref{assumption:for dpe} are satisfied.
 If $\ka \lo {\tilde X}$ and $\ka \lo {\tilde Y}$ are Gaussian radial basis kernels:
 $\ka_{\tilde X}(f_1, f_2) = \exp(-\gamma_X \|f_1-f_2\|^2_{\lx})$, and   $\ka_{\tilde Y}(f_1, f_2) = \exp(-\gamma_Y \|f_1-f_2\|^2_{\ly})$,
where  $\gamma_X > 0$, $ \gamma_Y >0$,
 then  $\tilde X \indep \tilde Y$ if and only if
 $\sigt=0$.
\end{corollary}

\def\hi{^}

\subsection{DPE-Based Test Statistic}

Having shown that the independence of the random objects $(X,Y)$ is equivalent to the vanishing of the cross-covariance operator between their DPEs, we now develop a corresponding DPE-based test of independence. By Theorem \ref{thm: equiv of idp}, it boils down to testing the hypothesis
\begin{equation}
    H_0: \sigt = 0.
\end{equation}
Let $(X \lo 1, Y \lo 1), \ldots, (X \lo n, Y \lo n)$ be  independent and identically distributed observations on $(X, Y)$,   let $(\tilde X \lo 1,  \tilde Y \lo 1), \ldots, (\tilde X \lo n, \tilde Y\lo n)$ be their DPEs, and let $\E \lo n$ denote the sample average; that is, $\E \lo n H(\tilde X, \tilde Y) = n \inv \sum \lo {i=1} \hi n H (\tilde X \lo i, \tilde Y  \lo i)$ for any function $H$. We estimate the cross-covariance operator $\sigt$ by
\begin{equation}\label{eq: hsigt}
    \hsigt = \E_n[\ka_{\tilde Y}(\dcd \tilde Y) \otimes \ka_{\tilde X}(\dcd \tilde X) ] - \E_n[\ka_{\tilde Y}(\dcd \tilde Y)] \otimes \E_n[\ka_{\tilde X}(\dcd \tilde X)].
\end{equation}

Let $\c B_{\HS}(\c H_{\tilde X},\c H_{\tilde Y})$ denote the space of Hilbert--Schmidt operators from
$\c H_{\tilde X}$ to $\c H_{\tilde Y}$. For a linear operator $T:\c H_{\tilde X}\to\c H_{\tilde Y}$, the Hilbert--Schmidt norm is defined as $\|T\|_{\HS}^2=\sum_{k\ge1}\|T e_k\|_{\c H_{\tilde Y}}^2,$
where $\{e_k\}$ is any orthonormal basis of $\c H_{\tilde X}$. We denote the tensor product on $\c B_{\HS}(\c H_{\tilde X},\c H_{\tilde Y})$ by $\otimes_{\HS}$.  The following theorem establishes the Central Limit Theorem (CLT) for the empirical cross-covariance operator $\hsigt$.
\begin{theorem}\label{thm: CLT UI}
Suppose that Assumptions \ref{assumption:d in L2 several} and \ref{assumption:for dpe} are satisfied, and the kernels $\ka_{\tilde X}$ and $\ka_{\tilde Y}$ are bounded. Then, as $n \to \infty$:
   \begin{equation}
    \sqrt{n} (\hsigt - \sigt) \cD \c N(0, \Gyx),
    \end{equation}
    where $\Gyx: \c B_{\HS}(\c H_{\tilde X},\c H_{\tilde Y}) \to \c B_{\HS}(\c H_{\tilde X},\c H_{\tilde Y})$ is the covariance operator defined by $\Gyx =  \E_{ \tilde X \tilde Y}[\{C_{\tilde Y \tilde X} - \sigt\}\otimes\HS \{C_{\tilde Y \tilde X} - \sigt\}]$ with $C_{\tilde Y \tilde X} = \{\ka_{\tilde Y}(\dcd\tilde Y) - \E_{\tilde Y}[\ka_{\tilde Y}(\dcd\tilde Y)]\} \otimes \{\ka_{\tilde X}(\dcd\tilde X) - \E_{\tilde X} \ka_{\tilde X}(\dcd\tilde X) \}$.
\end{theorem}
As shown in Theorem \ref{thm: CLT UI}, the proposed DPE-based framework does not require the structural assumptions typically required for the uniformly consistent estimation of profile-based distribution functions. Existing distance-profile-based approaches \citep{wang_nonparametric_2024, chen2024testing, zhou2025association} depend on the total boundedness and additional regularity conditions for the underlying distribution functions, such as finite entropy, covering number constraints, or Lipschitz-type assumptions, to ensure the convergence of the associated empirical processes. In contrast, the proposed framework does not treat distance profiles as intermediate nonparametric objects to be estimated pointwise. Instead, it operates directly through their global embedding to obtain a single element of a Hilbert space, thereby avoiding the restrictive regularity conditions inherent in the nonparametric estimation of univariate distributions.

The Central Limit Theorem established in Theorem~\ref{thm: CLT UI} motivates our proposed test statistic, defined as the squared Hilbert--Schmidt norm of the empirical cross-covariance operator:
\begin{equation}
T_n = n\|\hsigt\|^2\HS.
\end{equation}
To simplify notation, we define the empirically centered kernels $\hat\rho_{\tilde X}(\cdot,\tilde X_i)
:= \ka_{\tilde X}(\cdot,\tilde X_i)-\E_n\{\ka_{\tilde X}(\cdot,\tilde X)\}$ and
$\hat\rho_{\tilde Y}(\cdot,\tilde Y_i)
:= \ka_{\tilde Y}(\cdot,\tilde Y_i)-\E_n\{\ka_{\tilde Y}(\cdot,\tilde Y)\}$. The empirical operator can then be expressed as $\hsigt = \E_n[\hat\rho_{\tilde Y}(\cdot,\tilde Y_i)\otimes \hat\rho_{\tilde X}(\cdot,\tilde X_i)],$ which implies
\begin{align*}
    \|\hsigt\|_{\HS}^2
    = \langle \hsigt,\hsigt\rangle_{\HS} = \frac{1}{n^2}\sum_{i=1}^n\sum_{j=1}^n
\langle \hat\rho_{\tilde X}(\cdot,\tilde X_i),\hat\rho_{\tilde X}(\cdot,\tilde X_j)\rangle_{\c H_{\tilde X}}\,
\langle \hat\rho_{\tilde Y}(\cdot,\tilde Y_i),\hat\rho_{\tilde Y}(\cdot,\tilde Y_j)\rangle_{\c H_{\tilde Y}}.
\end{align*}

\subsection{Asymptotic Properties} In this subsection, we characterize the asymptotic behavior of the proposed test statistic under both the null and alternative hypotheses.
In addition to relaxing the conditions, our DPE framework provides an analytically tractable null distribution, which serves as a computationally efficient alternative to permutation-based resampling.

\def\nat{\mathbb{N}}

We begin by introducing two more empirical covariance operators:
\begin{align*}
\hSigtx  = \E_n[\hat \rho_{\tilde X}(\dcd \tilde X) \otimes \hat \rho_{\tilde X}(\dcd \tilde X) ], \quad \text{and} \quad
\hSigty  = \E_n[\hat \rho_{\tilde Y}(\dcd \tilde Y) \otimes \hat \rho_{\tilde Y}(\dcd \tilde Y) ].
\end{align*}
These operators are needed for characterizing the asymptotic null distribution of $T \lo n$.
The next theorem gives the asymptotic null distribution of $T \lo n$ and an approximation formula for its quantiles. In the following, we denote the $j$-th eigenvalue of an operator $T$ by $\lambda_j(T)$, and denote the set of natural numbers $\{1, 2, \ldots \}$ by $\nat$.

\begin{theorem}\label{thm: null dist}
 Under the assumptions of Theorem~\ref{thm: CLT UI}, let $\{W_{ij}: i,j \in \mathbb N\}$ be \emph{i.i.d.} standard normal random variables. Then, under the null hypothesis $H_0: \sigt = 0$,

     \begin{enumerate}
    \item [(i)] $T_n \cD \sum_{j=1}^\infty \lambda_j(\Gyx) W_{1, j}^2$, as $n\to \infty$;
    \item [(ii)] For every $t\in \R$, $P(T_n  \le t)  - P\bigl(\textstyle\sum_{i,j=1}^{n}\lambda_i(\hSigtx)\lambda_j(\hSigty) W_{i, j}^2 \le t\bigr) \rightarrow 0,$ as $n \to \infty$.
    \end{enumerate}
\end{theorem}
Theorem \ref{thm: null dist} provides the explicit asymptotic null distribution of the test statistic $T \lo n$ and an empirical approximation formula, building on the CLT for the estimated RKHS covariance operator from Theorem \ref{thm: CLT UI}. This marks a direct theoretical advance over the kernel methods \cite{gretton2007kernel, zhang2012kernel}, whose analysis was confined to scalar statistics. By directly characterizing the RKHS operators, our framework simultaneously provides the approximation tool, which is not available in \cite{gretton2007kernel}, and eliminates the need to truncate infinite summations in the null distribution as required by \cite{zhang2012kernel}.

While Theorem \ref{thm: null dist} ensures the asymptotic validity of the test, Theorem \ref{thm: alt dist} follows from Theorem~\ref{thm: CLT UI} and characterizes the asymptotic behavior of the test statistic under a fixed alternative.
\begin{theorem}\label{thm: alt dist}
Under the assumptions of Theorem~\ref{thm: CLT UI}, if $H_1$ is a fixed alternative with $\sigt \neq 0$, then
$$
    \sqrt{n}\left(\|\hsigt\|\HS^2 - \|\sigt\|\HS^2\right) \cD   N(0, 4\langle \Gyx \sigt, \sigt\rangle\HS).
$$
\end{theorem}
Theorem \ref{thm: alt dist} indicates that, under any fixe alternative $\sigt \neq 0$,  our test statistic $T_n \cP \infty$, confirming the consistency of the DPE-based independence test in the presence of any non-vanishing dependence between the random objects $X$ and $Y$.

Motivated by Theorem~\ref{thm: null dist}(ii), define the surrogate variable
\[
\hat R_n := \sum_{i,j=1}^{n}
\lambda_i(\hSigtx)\lambda_j(\hSigty)W_{ij}^2,
\]
and let $q_{1-\alpha}(\hat R_n)$ denote its $(1-\alpha)$-quantile. We reject the null at significance level $\alpha$ if $T_n > q_{1-\alpha}(\hat R_n)$. The next theorem further characterizes the sensitivity of our test statistic to the departure from the null hypothesis by examining its behavior under local alternatives. Specifically, we establish the limiting distribution of our test statistic when the sequence of local alternative hypotheses approaches the null hypothesis at a contiguous rate as the sample size goes to infinity.

\begin{theorem}\label{thm: power}
Under the assumptions of Theorem~\ref{thm: CLT UI}, consider the sequence $H^{(n)}_1: \sigt = a_n \Sigma_a$, where $\|\Sigma_a\|\HS < \infty$. Then:
    \begin{enumerate}
    \item [(i)] If $a_n n^{1/2} \rightarrow \infty$, the test statistic $T_n \cP \infty$ and the power $P(T_n > q_{1-\alpha}(\hat R_n)) \rightarrow 1$.
    \item [(ii)] Suppose $a_n = n^{-1/2}$,  $\Gyx=\sum_{j=1}^\infty \lambda_j(v_j\otimes v_j)$ for some orthonormal basis $\{v_1, v_2,\dots\}$ in $\c B_{\HS}(\c H_{\tilde X},\c H_{\tilde Y})$, and $\Sigma_a=\sum_{j=1}^\infty \sigma_jv_j$ with $\|\Sigma_a\|\HS < \infty$. Then,
    \(
    T_n \cD \sum_{j=1}^\infty\lambda_j \tilde W_j^2,
    \)
    where $\tilde W_j$ are independently distributed as $\mathcal{N}(\sigma_j/\sqrt{\lambda_j},1)$.
    \end{enumerate}
\end{theorem}

Among the current literature, only \cite{chen2024testing} and \cite{zhou2025association} provided asymptotic consistency results for independence testing of random objects. {Although \cite{chen2024testing} established consistency under shrinking alternatives and minimax rate optimality, while \cite{zhou2025association} derived a minimax separation bound, neither work derives explicit asymptotic distributions of the test statistics under such alternatives.}
In contrast, Theorem~\ref{thm: power} provides a more specific characterization of the test statistic under local alternatives, establishing an explicit noncentral chi-squared limiting distribution. This result offers new insights into the local power and asymptotic efficiency of distance-profile-based tests that have thus far been absent from the existing literature.

\subsection{Numerical Implementation}\label{subsection:numerical implementation}
Let $K_{\tilde X}\in \R^{n\times n}$ be the Gram matrix with entries $(K_{\tilde X})_{ij} = \ka_{\tilde X}(\tilde X_i, \tilde X_j)$, and let $Q = I_n - \frac 1 n 1_n 1_n^T$ denote the centering matrix, where $1_n$ is the $n$-dimensional vector of ones. This is the orthogonal projection onto the orthogonal complement of the linear subspace spanned by $1 \lo n$. By the reproducing property of the RKHS, the inner product of the centered kernels satisfies $\langle \hat\rho_{\tilde X}(\cdot,\tilde X_i),\hat\rho_{\tilde X}(\cdot,\tilde X_j)\rangle_{\c H_{\tilde X}}
= (QK_{\tilde X}Q)_{ij},$
and a similar result holds for $Y$. The squared Hilbert-Schmidt norm of $\hat \Sigma \lo {\tilde X \tilde Y}$ admits the following trace representation:
$$
\|\hsigt\|_{\HS}^2 =\frac{1}{n^2}\sum_{i,j=1}^n (QK_{\tilde X}Q)_{ij} (QK_{\tilde Y}Q)_{ij}
=\frac{1}{n^2}\tr\!\big(K_{\tilde X}QK_{\tilde Y}Q\big),
$$
where we used
$Q^\top=Q$ and $Q^2=Q$. Thus, the test statistic $T_n$ admits the explicit form:
\begin{equation}\label{eq: Tn1}
T_n
= \frac{1}{n}\tr\!\big(K_{\tilde X}QK_{\tilde Y}Q\big).
\end{equation}
This formulation identifies $T_n$ as the empirical Hilbert--Schmidt Independence Criterion (hSIC) statistic \citep{gretton2007kernel}. While hSIC is utilized for Euclidean data, our framework extends its utility to complex random objects by embedding their metric structures via the DPE.

To compute the surrogate variable $\hat R_n$, we obtain
$\lambda_j(\hSigtx)$ using the coordinate mappings for $\hSigtx$ and the
notation of \citet{li2018sufficient}. Let $\c B_{\tilde X} = \{\ka_{\tilde X}(\dcd \tilde X_i) - \E_n[\ka_{\tilde X}(\dcd \tilde X)]: i=1,\dots n\}$ as a basis of the linear space $\ran(\hSigtx)$. According to \citet{li2018sufficient}, the coordinate representation is
\(
_{\c B_{\tilde X}}[\hSigtx]_{\c B_{\tilde X}} = n^{-1}QK_{\tilde X}Q.
\)
Thus, we obtain $\lambda_j(\hSigtx)$ by computing the eigenvalues of $n^{-1}QK_{\tilde X}Q$. This applies to $\hSigty$ analogously.
Following \cite{gretton2007kernel} and \cite{zhang2012kernel}, we can also utilize a Gamma approximation to provide a computationally efficient estimate of $q_{1-\alpha}(\hat R_n)$ and compute the associated $p$-value.

\section{DPE-Based Conditional Independence Testing}\label{sec: CI}
This section extends our DPE framework to the problem of testing conditional independence among three random objects. Again, our method does not require any structural assumptions on the underlying metric spaces, such as the negative-type property, mirroring the flexibility achieved in Section \ref{sec: UI}.

\subsection{DPE-Based Characterization of Conditional Independence}

{ As in Subsection \ref{subsection:preserving},
let $(X, Y, Z)$ be a random object taking values in the product measurable space
$( \Omega \lo X \times \Omega \lo Y \times \Omega \lo Z, \ \ca F \lo X \times \ca F \lo Y \times \ca F \lo Z)$, and let $(\tilde X, \tilde Y, \tilde Z)$ be the DPE of $(X, Y, Z)$. By Corollary \ref{corollary:conditional independence}, testing $X\indep Y | Z$ is equivalent to testing $\tilde X \indep \tilde Y | \tilde Z$, where all random elements involved are Hilbertian.
The key idea behind testing the conditional independence of two Hilbertian random elements given a third is to regress out the effect of the conditioning variable and then assess whether any dependence remains. This remains the guiding principle in our setting as well.

\def\real{\mathbb{R}}

Following the notation in Section \ref{subsection:measurability}, we use $\Phi \lo {XZ}(X,Z)$ to represent the DPE of $(X, Z)$ and abbreviate it by $(\tilde X, \tilde Z)$.
Let $\ka_{\tilde Z}: L \lo 2 (\Omega \lo Z, \lambda \lo Z) \times L \lo 2 (\Omega \lo Z, \lambda \lo Z) \to \real $ be a positive definite kernel. }

{
\def\ali{&\ \,  }
{We  construct the product kernel  $\ka_{\tilde X \tilde Z}=\ka_{\tilde X}\ka_{\tilde Z}$, which is the map
\begin{align*}
\ka \lo {\tilde X \tilde Z} : \ali   [  L \lo 2 (\Omega \lo X, \lambda \lo X) \times   L \lo 2 (\Omega \lo Z, \lambda \lo Z)] \times
[  L \lo 2 (\Omega \lo X, \lambda \lo X) \times   L \lo 2 (\Omega \lo Z, \lambda \lo Z)] \to \real.
\end{align*}
Let $\c H_{\tilde X \tilde Z}, \c H_{\tilde Y}$, and $\c H_{\tilde Z}$ denote the RKHSs generated by $\ka_{\tilde X \tilde Z}, \ka_{\tilde Y }$, and $\ka_{\tilde Z}$, respectively. Let $\Sigma \lo { \tilde Y    (\tilde X \tilde Z) }:   \ca H \lo { \tilde X \tilde Z} \to  \ca H \lo { \tilde Y},$ $\Sigma \lo {(\tilde X \tilde Z)  \tilde Z  }:   \ca H \lo {  \tilde Z} \to  \ca H \lo { \tilde X \tilde Z}$ and $\Sigma \lo { \tilde Y    \tilde Z  }:   \ca H \lo {  \tilde Z} \to  \ca H \lo { \tilde Y }$
be the covariance operators defined according to the same rule used in (\ref{eq:Sigma XY}). For example,
\begin{align*}
     \Sigma \lo { \tilde Y    (\tilde X \tilde Z) }
     = \ali
     \E [\ka \lo {  \tilde Y  }( \cdot,  \tilde Y ) \otimes
    \ka \lo {  \tilde X \tilde Z}( \cdot, (\tilde X, \tilde Z))] -
     \E [\ka \lo {  \tilde Y  }( \cdot,  \tilde Y )
   ] \otimes
  \E [
    \ka \lo {  \tilde X \tilde Z}( \cdot, (\tilde X, \tilde Z))].
\end{align*}

To construct the desired operator for testing $\tilde X \indep \tilde Y | \tilde Z$, we need to introduce the Moore-Penrose inverse and a related range condition. For a linear operator $A: \ca H \to \ca K$ between Hilbert spaces  $\ca H$ and $\ca K$, we let $\ran (A)$ denote the range of $A$: $\ran (A) = \{A f: f \in \ca H\}$, let $\cran (A)$ denote the closure of $\ran(A)$, let $\ker (A)$ denote the kernel of $A$: $\ker (A) = \{ f \in \ca H: Af = 0 \}$. For a self-adjoint operator $A: \ca H \to \ca H$, $\cran(A) = \ker(A) \hi \perp$. If $\tilde A = A | \cran (A)$ is $A$ restricted to $\cran (A)$, then $\tilde A$ is an injection, and its inverse mapping $\tilde A \inv$ is defined. We call this inverse the Moore-Penrose inverse of $A$, and denote it by $A \hi \dagger$. See \cite{hsing2015theoretical} and  \cite{li2018sufficient}. Since the domain of $A \hi \dagger$ is $\ran (A)$, for an operator of the form $A \hi \dagger B$ to be defined, we need to assume $\ran(B) \subseteq \ran (A)$. For this reason, we make the following assumption.
\begin{assumption}\label{assumption:ran in ran}
    $\ran ( \Sigma \lo {\tilde Z( \tilde X \tilde Z)} ) \subseteq \ran ( \Sigma \lo {\tilde Z \tilde Z} )$, and  $\ran ( \Sigma \lo {\tilde Z  \tilde Y } ) \subseteq \ran ( \Sigma \lo {\tilde Z \tilde Z} )$.
\end{assumption}
Under this assumption, the operators $\Sigma \lo {\tilde Z \tilde Z} \hi \dagger \Sigma \lo {\tilde Z ( \tilde X \tilde Z)}$ and $\Sigma \lo {\tilde Z \tilde Z} \hi \dagger \Sigma \lo {\tilde Z   \tilde Y }$ are defined, and therefore the following operator is also defined:
\begin{equation}\label{eq: Sigxyz}
    \Sigxyz = \Sigma_{ \tilde Y  (\tilde X \tilde Z)}
    - \Sigma_{ \tilde Y  \tilde Z}\Sigma_{\tilde Z\tilde Z}^\dagger \Sigma_{\tilde Z(\tilde X \tilde Z)},
\end{equation}
This operator was introduced by \cite{fukumizu2007kernel}. Intuitively,
it captures the residual dependence between $(\tilde X, \tilde Z)$ and $\tilde Y$ after removing the effect of $\tilde Z$ by Hilbert-space regression.

\def\cov{\mathrm{cov}}

To provide a rigorous characterization of conditional independence, we introduce the following assumption regarding the representability of the conditional expectation.

\begin{assumption}\label{as: GZ}
    For every $f\in \c H_{\tilde X \tilde Z}$ and $ g\in \c H_{\tilde Y }$, the conditional expectations satisfy that $\E[f(\tilde X, \tilde Z )| \tilde Z ] \in \c H_{\tilde Z}$, and $\E[g(\tilde Y  )|\tilde Z ] \in \c H_{\tilde Z}$.
\end{assumption}
This assumption is to guarantee that conditional expectations are contained within the RKHS of the conditioning variable $\tilde Z$, enabling the use of kernel-based projection to isolate the residual components of variation. This is a standard condition in the theory of kernel-based conditional dependence tests \citep{fukumizu2004dimensionality}. Under this assumption, by \cite{fukumizu2004dimensionality}, it can be shown that the conjoined conditional covariance operator satisfies the first equality below:
$$
\langle g, \Sigxyz f\rangle_{\c H_{\tilde Y}} = \E_{\tilde Z} [\cov (f(\tilde X, \tilde Z ), g(\tilde Y)| \tilde Z )  ]= \E_{ Z} [\cov (f(\tilde  X, \tilde Z), g(\tilde Y)| Z)  ],
$$
where the second equality holds because $Z$ and $\tilde Z$ generates the same sub-$\sigma$-field in $\ca F$.
We now establish that the vanishing of this operator, i.e., $\Sigxyz=0$,  is equivalent to conditional independence when characteristic kernels are employed. Let $P_{\tilde Z}:=P\circ\tilde Z^{-1}$ denote the distribution of $\tilde Z$ on $L \lo 2(\Omega_Z,\lambda_Z)$. In what follows, let $L \lo 2 (L \lo 2 (\Omega \lo Z, \lambda \lo Z), P_{\tilde Z})=   \{f: L \lo 2 (\Omega \lo Z, \lambda \lo Z) \rightarrow \R,\ \E_{\tilde Z}[f^2(\tilde Z)] < \infty\}$  denote the collection of all real-valued functions defined on $L \lo 2 (\Omega \lo Z,\lambda \lo Z)$ that are square-integrable with respect to $P_{\tilde Z}$.
This notation reflects the underlying embedding structure of our method.

The next theorem establishes a general criterion for conditional independence that requires no Euclidean structure, relying on a characteristic product kernel defined on the DPE spaces.

\begin{theorem}\label{thm: equiv of con idp}
Suppose Assumptions \ref{assumption:d in L2 several}, \ref{assumption:for dpe}, \ref{assumption:ran in ran}, and \ref{as: GZ} are satisfied and the kernels $\ka_{\tilde X}$, $\ka_{\tilde Y}$, and $\ka_{\tilde Z}$ are bounded. Assume that the product kernel $\ka_{\tilde X} \ka_{\tilde Y} \ka_{\tilde Z}$ is characteristic, and that
$\c H_{\tilde Z} + \R$ is dense in
$L \lo 2(L \lo 2(\Omega_Z,\lambda_Z), P_{\tilde Z})$.
Then
\begin{equation}\label{eq: sig and CI}
    \Sigxyz =  0 \, \Leftrightarrow \, \tilde X\indep \tilde Y | \tilde Z.
\end{equation}
\end{theorem}

Gaussian radial basis kernels satisfy the characteristic-product and density conditions in Theorem \ref{thm: equiv of con idp}, as shown in the Supplementary Material. We record this important special case in the next corollary.

\begin{corollary}\label{cor: RBF con}
If Assumptions \ref{assumption:d in L2 several}, \ref{assumption:for dpe}, \ref{assumption:ran in ran}, and \ref{as: GZ} are satisfied, and   $\ka_{\tilde X}, \ka_{\tilde Y}, \ka_{\tilde Z}$ are Gaussian radial basis kernels, then equivalence (\ref{eq: sig and CI}) holds.
\end{corollary}
Theorem \ref{thm: equiv of con idp} and Corollary \ref{cor: RBF con}, along with Corollary \ref{corollary:conditional independence}, establish the equivalence between conditional independence $X \indep Y |Z$ and the vanishing of the conditional cross-covariance operator $\Sigxyz$. This equivalence provides a general criterion for testing the conditional independence of random objects that relies on a characteristic product kernel and on mild conditions on the reference measures and metric spaces involved. It constitutes the novel contribution of our approach as highlighted in Table \ref{tab:conditional_method_comparison}.
In contrast to the PA test by \citet{zhou2025association}, which restricts the conditioning variable to Euclidean spaces, and the KCI framework by \citet{zhang2012kernel}, which requires Hilbert structures for all variables, our approach does not impose any geometric restrictions on the conditioning random object.

\subsection{DPE-Based Test Statistic and its Asymptotic Properties}

To construct a test statistic for conditional independence, we first write the conditional cross-covariance operator in \eqref{eq: Sigxyz} equivalently as
\begin{equation}\label{eq: Sigxyz_para}
    \Sigxyz
    = \Sigma_{\tilde Y(\tilde X\tilde Z)}
    {-2\Sigma_{\tilde Y\tilde Z}\Sigz^\dagger
      \Sigma_{\tilde Z(\tilde X\tilde Z)}
    +\Sigma_{\tilde Y\tilde Z}\Sigz^\dagger
      \Sigma_{\tilde Z\tilde Z}\Sigz^\dagger
      \Sigma_{\tilde Z(\tilde X\tilde Z)}.}
\end{equation}
Since $\Sigma \lo {\tilde Z \tilde Z}$ is a Hilbert-Schmidt operator when it is defined (see, for example, \cite{li2018linear}), $\Sigma \lo {\tilde Z \tilde Z} \hi \dagger$ is an unbounded operator and cannot be estimated consistently. However, as argued in \cite{li2018linear}, it is reasonable assume $\Sigz^\dagger \Sigma_{\tilde Z (\tilde X \tilde Z)}$ and $\Sigma_{ \tilde Y  \tilde Z}\Sigz^\dagger $ to be bounded. In fact, to ensure an appropriate convergence rate, we impose stronger conditions on these two operators.
Define the population-centered feature map by $\rho_{\tilde X}(\cdot,\tilde x)=\ka_{\tilde X}(\cdot,\tilde x)-\E\{\ka_{\tilde X}(\cdot,\tilde X)\}$, and define $\rho_{\tilde Y}$, $\rho_{\tilde Z}$, and $\rho_{(\tilde X\tilde Z)}$ analogously. Let $\Lambda_{(\tilde X \tilde Z) \tilde Z} = \E[{\rho}_{\tilde X}(\cdot,\tilde X)\kappa_{\tilde Z}(\cdot,\tilde Z) \otimes {\rho}_{\tilde Z}(\cdot,\tilde Z)]$.
This operator has an additional layer of centering compared with $\Sigma \lo {(\tilde X \tilde Z) \tilde Z}$. Let $\mu_{\tilde X}$ be the mean element $\E [\ka \lo {\tilde X} ( \cdot, \tilde X)]$. Then $\Sigma_{(\tilde X \tilde Z) \tilde Z} =\mu_{\tilde X} \Sigma_{\tilde Z \tilde Z} + \Lambda_{(\tilde X \tilde Z ) \tilde Z}$,
where $\mu_{\tilde X}\Sigma_{\tilde Z\tilde Z}: \ca H \lo {\tilde Z} \to \ca H \lo {\tilde X \tilde Z}$ is the linear operator defined by $( \mu_{\tilde X} \Sigma \lo {\tilde Z \tilde Z} ) f = \mu_{\tilde X} (\Sigma \lo {\tilde Z \tilde Z} f)$ for any $f \in \ca H \lo {\tilde Z}$.
We impose the following assumption on
$\Lambda_{(\tilde X\tilde Z)\tilde Z}$ and
$\Sigma_{\tilde Y\tilde Z}$.

\begin{assumption}\label{as: S}
There exist bounded linear operators $S_{\tilde Y\tilde Z}:\c H_{\tilde Z}\to\c H_{\tilde Y}$ and
{$S_{(\tilde X\tilde Z)\tilde Z}:
\c H_{\tilde Z}\to\c H_{\tilde X\tilde Z}$}, and
$\beta>0$, such that
{\(\Sigma_{\tilde Y\tilde Z}
 =S_{\tilde Y\tilde Z}\Sigma_{\tilde Z\tilde Z}^{1+\beta}\)}, and
{\(\Lambda_{(\tilde X\tilde Z)\tilde Z}
=S_{(\tilde X\tilde Z)\tilde Z}
\Sigma_{\tilde Z\tilde Z}^{1+\beta}.
\)}
\end{assumption}
This is essentially a smoothness condition. For more insights and concrete examples about this assumption, see \cite{tang_CCCO_2026}, \cite{bing2017SDR}, \cite{bhattacharjeeNonlinearGlobalFrechet2023}, and \cite{sang_nonlinear_2022}. For two nonnegative sequences $\{a_j\}$ and $\{b_j\}$, we write {$a_j\preceq b_j$} if there exists a constant $C>0$ such that $a_j\le Cb_j$ for all $j$. We make the following assumption about $\Sigma \lo {\tilde Z \tilde Z}$.
\begin{assumption}\label{as: lams of Sigz}
The eigenvalues of $\Sigma_{\tilde Z \tilde Z}$ satisfy
$\lambda_j(\Sigma_{\tilde Z \tilde Z}) \preceq j^{-\eta}$
for some $\eta > 1$.
\end{assumption}

The equivalent formulation \eqref{eq: Sigxyz_para} motivates a plug-in estimator obtained by replacing the population covariance operators with their empirical counterparts. To estimate the Moore--Penrose pseudoinverse $\Sigz^\dagger$, we employ Tikhonov regularization. Let the empirical version of the cross-covariance operators be defined in analogy to $\hsigt$, and $\hat\Sigma_{\tilde Z\tilde Z}^{\dagger}=(\hat\Sigma_{\tilde Z\tilde Z} + \epsn I)^{-1}$ is the regularized inverse with a tuning constant $\epsn > 0$. We introduce the plug-in estimator:
\begin{equation}\label{eq: hSigxyz}
        \hSigxyz = \hSigdyx - 2\hSigdyz\hSigzd \hSigdzx + \hSigdyz\hSigzd \hSigz\hSigzd\hSigdzx.
\end{equation}
 Let $a \wedge b := \min(a, b)$, and let
$\mathcal B \HS (\mathcal H_{ \tilde X \tilde Z }, \mathcal H_{ \tilde Y })$ denote the space of Hilbert-Schmidt operators from $\mathcal H_{ \tilde X \tilde Z }$ to $\mathcal H_{ \tilde Y }$. We write $a_n \asymp b_n$ if $a_n/b_n$ converges to a positive constant as $n \to \infty$.

Let $C_{  \tilde Y  (\tilde X \tilde Z)| \tilde Z} =  D_{ \tilde Y   \tilde Z}   \otimes D_{(\tilde X \tilde Z) \tilde Z},$ where $D_{ \tilde Y   \tilde Z} =  \rho_{ \tilde Y  }(\dcd \tilde Y) - \Sigma_{ \tilde Y  \tilde Z} \Sigma_{\tilde Z \tilde Z}^\dagger \rho_{\tilde Z}(\dcd \tilde Z)$, $D_{(\tilde X \tilde Z) \tilde Z} =  \rho_{(\tilde X \tilde Z)}(\dcd (\tilde X,\tilde Z)) - \Sigma_{(\tilde X \tilde Z)\tilde Z} \Sigma_{\tilde Z \tilde Z}^\dagger \rho_{\tilde Z}(\dcd \tilde Z)$. Now, define $B_{ \tilde Y   (\tilde X \tilde Z)| \tilde Z} =  C _{ \tilde Y   (\tilde X \tilde Z)| \tilde Z} - \Sigma _{ \tilde Y   (\tilde X \tilde Z)| \tilde Z},$ and $\Gamma_{  \tilde Y  (\tilde X \tilde Z)| \tilde Z} = \E( B_{ \tilde Y  (\tilde X \tilde Z)| \tilde Z}  \otimes    B_{  \tilde Y  (\tilde X \tilde Z)| \tilde Z}).$

Under a suitable convergence rate of $\epsilon \lo n$, we next establish the asymptotic normality of the plug-in estimator $\hSigxyz$ in \eqref{eq: hSigxyz} in the following theorem.

\begin{theorem}\label{thm: CLT CI}
Suppose that Assumptions \ref{assumption:d in L2 several}--\ref{assumption:ran in ran} are satisfied and \ref{as: GZ}--\ref{as: lams of Sigz} hold  for some   $\beta,\eta$  satisfying $\beta>\frac{\eta-1}{2 \eta}$ and $\frac{\eta(\beta \wedge 1)}{2 \eta(\beta \wedge 1)+\eta+1}>\frac{1}{4}$. Suppose also that the kernels $\ka_{\tilde X}$, $\ka_{\tilde Y}$, and $\ka_{\tilde Z}$ are bounded. If, moreover, \  $\epsn \asymp n^{-\frac{\eta(\beta \wedge 1)}{2 \eta(\beta \wedge 1)+\eta+1}}$, \ then
\begin{equation}
    \sqrt{n}\left(\hSigxyz-\Sigxyz\right) \xrightarrow{\mathcal{D}} N\left(0, \Gamma_{  \tilde Y   (\tilde X \tilde Z)| \tilde Z}\right).
\end{equation}
\end{theorem}
Theorem \ref{thm: CLT CI} establishes a Central Limit Theorem for the empirical conditional cross-covariance operator. While previous work, such as \cite{fukumizu2007kernel} and \cite{liu2025sparse}, obtained the convergence rate for such operators, the explicit asymptotic normality has remained largely undeveloped. By characterizing the limiting behavior of the conditional operator, Theorem \ref{thm: CLT CI} provides the necessary theoretical framework to rigorously evaluate the validity and statistical power of the DPE-based conditional independence test.}}

Our test statistic for the conditional independence test is
\begin{equation}\label{eq: Sn}
S_n = n\|\hSigxyz\|^2\HS.
\end{equation}
Parallel to the unconditional independence test case, the above statistic can be represented as the trace of an $n \times n$ matrix.
Define the $n \times n$ Gram matrices for $\ka_{ \tilde X \tilde Z }, \ka_{ \tilde Y  }$ and $\ka_{\tilde Z}$: $K_{ \tilde X \tilde Z } =\{ \ka \lo { \tilde X \tilde Z } ( ( \tilde X \lo i,  \tilde Z \lo i ),( \tilde X \lo j,  \tilde Z \lo j )): i,j =1, \ldots, n \},$ $K_{ \tilde Y   }= \{ \ka \lo { \tilde Y } (   \tilde Y \lo i ,  \tilde Y \lo j ): i,j =1, \ldots, n \},$ $K_{\tilde Z}=\{ \ka \lo {  \tilde Z } (    \tilde Z \lo i,  \tilde Z \lo j ): i,j =1, \ldots, n \},$
and their centered versions: $G_{ \tilde X \tilde Z } = Q K_{ \tilde X \tilde Z } Q$, $G_{ \tilde Y  } = QK_{ \tilde Y } Q$, $G_{\tilde Z} = Q K_{  \tilde Z } Q$,
where $Q$ is as defined in Section \ref{subsection:numerical implementation}.
Let $R_{\tilde Z} = I-G_{\tilde Z}(G_{\tilde Z}+n\epsn I)^{-1}$, and then let $\Gx =  R_{\tilde Z}G_{ \tilde X \tilde Z }R_{\tilde Z}$ and $\Gy=R_{\tilde Z} G_{ \tilde Y   } R_{\tilde Z}$.
As shown in Lemma S3 of the Supplementary Material,  $S_n$ can be rewritten as the following trace of a matrix:
$$
S_n  = n \inv \, \tr(\Gx\Gy).
$$
The statistic $S_n$ preserves the computational structure of the DPE-based independence test \eqref{eq: Tn1}, highlighting the intrinsic structural simplicity of our DPE framework.

Leveraging the central limit theorem established in Theorem \ref{thm: CLT CI}, we derive the asymptotic null distribution of $S_n$ in Theorem \ref{thm: null dist CI}.

\begin{theorem}\label{thm: null dist CI}
Under the assumptions of Theorem~\ref{thm: CLT CI} and the null hypothesis $H_0: \Sigxyz = 0$, we have
$S_n \cD \sum_{j=1}^\infty \,  \lambda_j(\Gyxz) \,  W_{j}^2,$
where $\lambda_j(\Gyxz)$'s are the eigenvalues of the operator $\Gyxz$ in Theorem~\ref{thm: CLT CI}, and  $\{W \lo 1, W \lo 2, \ldots \}$ are  \emph{i.i.d.} standard normal variables.
\end{theorem}

In what follows, we provide an explicit form of the null distribution for the conditional independence test statistic $S_n$. Theorems \ref{thm: CLT CI} and \ref{thm: null dist CI} significantly advance the theoretical framework of \cite{zhang2012kernel}, which lacked operator-level asymptotic distribution and, as a result, did not provide a formal analytical benchmark distribution to perform local power analysis.

In practice, the eigenvalues $ \lambda_j(\Gyxz)$ in Theorem \ref{thm: null dist CI} are unknown. We construct a surrogate null distribution which provably approximates the true null following the Proposition 5 of \citet{zhang2012kernel}. Specifically, we approximate the null distribution of $S_n$ via the surrogate variable:
$$
\ddot R_n := n^{-1}\textstyle \sum_{j=1}^n \lambda_j(G_{\tilde X \tilde Z| \tilde Z} \odot \Gy) W_{j}^2,
$$
where $G_{\tilde X \tilde Z| \tilde Z} \odot \Gy \in \real \hi {n \times n}$ is the Hadamard product matrix with its $(i,j)$-th entry being $ (\Gx)_{ij}(\Gy)_{ij}$, which is positive semidefinite by Schur's theorem.
Let $  q_{1-\alpha}(\ddot R_n)$ denote the $(1-\alpha)$-quantile of $\ddot R_n$. We reject the null hypothesis at significance level $\alpha$ if
$$S_n >   q_{1-\alpha}(\ddot R_n).$$
Compared to \cite{zhang2012kernel}, our approach reduces the computational burden by requiring only half as many eigen-decompositions, while preserving the exact distribution of the surrogate variable $\ddot R_n$, see Lemma~S4 of the Supplementary Material.  For practical implementation, the distribution of $\ddot R_n$ can be efficiently approximated using a Gamma distribution, as detailed in Section~S1.3 of the Supplementary Material.

Next, we establish the consistency of the DPE-based conditional independence test. The following result is a direct consequence of Theorem~\ref{thm: CLT CI} and characterizes the asymptotic behavior of the test statistic $S_n$ under a fixed alternative.
\begin{theorem}\label{thm: alt dist CI}
Under the assumptions of Theorem~\ref{thm: CLT CI} and a fixed alternative hypothesis with $\Sigxyz \neq 0$, we have:
\begin{align*}
     \sqrt{n}\left(\|\hSigxyz\|\HS^2 - \|\Sigxyz\|\HS^2\right) \cD \c N\left(0, 4\langle \Gyxz \Sigxyz, \Sigxyz\rangle\HS\right).
\end{align*}
\end{theorem}
Under a fixed alternative,  Theorem \ref{thm: alt dist CI} implies that $ S_n \cP \infty$, establishing the consistency of the proposed test. We next characterize the asymptotic properties of $S_n$ under local alternative distributions.
\begin{theorem}\label{thm: power CI}
    Under the assumptions of Theorem~\ref{thm: CLT CI}, consider the sequence of local alternatives $\Sigxyz = a_n \ddot\Sigma_a$ with $0<\|\ddot \Sigma_a\|\HS < \infty$. Then:
    \begin{enumerate}
    \item [(i)] If $a_n n^{1/2} \rightarrow \infty$, then $S_n \cP \infty$ and the power $P(S_n > q_{1-\alpha}(\ddot R_n)) \rightarrow 1$.

    \item [(ii)] Suppose  $a_n = n^{-1/2}$. Let $\Gyxz$ have  spectral decomposition $\sum_{j=1}^\infty  \lambda_j( v_j\otimes v_j)$, where $  v_1,   v_2,\dots$ is an orthonormal basis of $\c B_{\HS}(\c H_{\tilde X\tilde Z}, \c H_{\tilde Y})$ and assume  $ \ddot\Sigma_a=\sum_{j=1}^\infty   \sigma_j   v_j$. Then
    \(
    S_n \cD \sum_{j=1}^\infty\lambda_j \tilde W_j^2,
    \)
    where $\tilde W_j$ are independently distributed as $\mathcal{N}(\sigma_j\lambda_j^{-1/2},1)$.
    \end{enumerate}
\end{theorem}
As highlighted in Table \ref{tab:conditional_method_comparison}, existing conditional testing framework KCI \citep{zhang2012kernel} lack theoretical consistency guarantees, largely because of the intrinsic difficulty of modeling conditional structures in non-Euclidean spaces. {The conditional PA test \citep{zhou2025association} established rejection consistency for alternatives, but does not derive the explicit limiting distribution or local power function under boundary-rate alternatives.}
We address the limitations by explicitly deriving the asymptotic behavior of our statistic under both fixed and local alternatives. Theorems \ref{thm: alt dist CI} and \ref{thm: power CI} provide a rigorous characterization of the test’s sensitivity.

In summary, Section~\ref{sec: CI} addresses an important gap in the literature by establishing a comprehensive framework for DPE-based conditional independence testing. The proposed framework provides rigorous guarantees for validity, asymptotic convergence, and power when analyzing object-valued data in general metric spaces. These theoretical results underscore the fundamental role of DPE in studying conditional independence among  random objects in a general metric space without requiring any additional structure.

\section{Implementation and Choice of Reference Measure}\label{sec: referen measure}
One of the key components of DPE is the choice of the reference measures $\lambda_X$, $\lambda_Y$, and $\lambda_Z$, which, as shown in Sections~\ref{sec: UI} and \ref{sec: CI}, determine the geometry of the embedding. In this section, we propose several natural choices for these measures, with particular emphasis on metric spaces that do not admit an isometric Hilbert embedding, where the DPE framework is especially needed.
Specifically, as shown in Corollaries \ref{cor: RBF UI} and \ref{cor: RBF con}, constructing both the unconditional and conditional independence tests requires reference measures for the kernels of the RKHSs involved. Taking $X$ as an illustration, the kernel $\ka \lo {\tilde X}$ is constructed from the squared $L \lo 2 (\Omega \lo X, \lambda \lo X)$ norm
\begin{equation}\label{eq: Jx}
   \|\Phi_X(x_1)-\Phi_X(x_2)\|^2_{\lx} = \int_{u\in \Ox} [d_X(u, x_1) - d_X(u, x_2)]^2 d\lx(u).
\end{equation}
We now discuss several strategies for constructing $\lambda \lo X$ according to the nature of $\Omega \lo X$.
\begin{itemize}
    \item \emph{Explicit Integration via Spherical Harmonics.}
For certain structured triples $(\Omega_X,d_X,\lambda_X)$, the integral in \eqref{eq: Jx} admits an explicit representation. A representative example is the sphere $\mathbb{S}^{p-1}$ equipped with the geodesic distance
\(
d_g(x_1,x_2)=\arccos\langle x_1,x_2\rangle.
\)
As shown in Lemma~S2 of the Supplementary Material, $(\mathbb{S}^{p-1}, d_g)$ does not admit an isometric Hilbert embedding when $p\ge3$. Choosing $\lambda_X$ to be the normalized Euclidean surface measure on $\mathbb{S}^{p-1}$, so that the total surface measure is one, the integral in \eqref{eq: Jx} can be evaluated using spherical harmonics \citep{atkinson2012spherical}.
The key idea is to represent functions on the sphere using an orthogonal basis, analogous to the Fourier basis for periodic functions. This representation transforms the $\lambda_X$-inner product of geodesic distance functions into a weighted sum of simple kernel terms, yielding an exact and numerically stable implementation (see Section~S1.2 of the Supplementary Material for details).

    \item \emph{Monte Carlo Approximation.} A more general approach is to let $\lambda \lo X$ be a probability measure for $\lx$ and use Monte Carlo integration to approximate  \eqref{eq: Jx}. This is applicable when it is feasible to generate random samples from a distribution defined on the underlying metric space.
For example, the $p\times p$ SPD matrix space $\c S_{++}^p$ is commonly equipped with  Affine Invariant Riemannian Metric (AIRM), $d_R(A, B) = \|\log(A^{-1/2}BA^{-1/2})\|_F$, where $\|\cdot\|_F$ denotes the Frobenius norm. The metric has been widely used in medical imaging, continuum mechanics, radar signal processing, and computer vision \citep{said2017riemannian}. Since the space $(\c S_{++}^p, d_R)$ does not admit an isometric embedding into Hilbert space \citep{jayasumana2013kernel}, we can approximate \eqref{eq: Jx} by sampling from a fully supported probability measure on $\c S_{++}^p$, such as the Riemannian Gaussian distributions \citep{said2017riemannian} and Wishart distributions.

    \item \emph{The Empirical Reference Measure.} When evaluating the integral in \eqref{eq: Jx} is computationally demanding and sampling from a distribution on the metric space is infeasible, the empirical distribution based on $\{X_1,\dots,X_n\}$ provides a natural surrogate for $\lambda_X$. Specifically, we take
$
\lambda_X=\frac{1}{n}\sum_{i=1}^n\delta_{X_i},
$
where $\delta_a$ denotes the Dirac measure at $a$. Under this choice, \eqref{eq: Jx} reduces to the simple average
$n \inv \sum_{i=1}^n [d_X(X_i,x_1)-d_X(X_i,x_2) ]^2$.
In practice, this empirical reference measure is always available and straightforward to implement when the metric space lacks a tractable reference measure or when Monte Carlo integration is computationally prohibitive. Moreover, it preserves the characteristic properties of the DPE, provided the sample provides sufficient coverage of the underlying support.

\end{itemize}

\section{Numerical Experiments}\label{sec: numerics}

This section evaluates the empirical performance of our DPE-based framework for testing independence and conditional independence through simulations on three metric spaces that do not admit an isometric embedding into a Hilbert space.

\subsection{Metric spaces}
We consider the following three scenarios:
\begin{itemize}
\item \textbf{Scenario 1: Geodesic distances on the two-dimensional sphere $\mathbb{S}^2$.}

\item \textbf{Scenario 2: SPD matrices under AIRM.} The space of symmetric positive-definite (SPD) matrices $(\mathcal{S}_{++}^p, d_R)$ endowed with the AIRM $d_R(A, B) = \|\log(A^{-1/2}BA^{-1/2})\|_F$.

\item \textbf{Scenario 3: Bivariate Gaussian distributions under Wasserstein distance.} The space of bivariate Gaussian measures equipped with the Wasserstein--2 metric.
\end{itemize}
In all three scenarios, the underlying metric spaces do not admit isometric embeddings into Hilbert spaces. The non-embeddability of the first two scenarios was established in Section~\ref{sec: referen measure}.
For Scenario 3, consider two probability measures $\mu$ and $\nu$ on $\mathbb{R}^d$ with finite second moments. The Wasserstein--2 distance is defined as $W_2(\mu,\nu)
=
\left(
\inf_{\pi \in \Pi(\mu,\nu)}
\int_{\mathbb{R}^d \times \mathbb{R}^d}
\|x-y\|^2 \, d\pi(x,y)
\right)^{1/2},$
where $\Pi(\mu,\nu)$ is the set of all joint distributions of $(X,Y)$ with  $\mu$ and $\nu$ as marginals for $X$ and $Y$ \citep{zhang2024copula}.
In the special case of Gaussian distributions on $\mathbb{R}^d$, where
$\mu=N(m,\Sigma)$ and $\nu=N(m',\Sigma')$, the squared Wasserstein distance admits the closed-form expression
$W_2^2(\mu,\nu)
=
\|m-m'\|^2
+
\operatorname{tr}(\Sigma)
+
\operatorname{tr}(\Sigma')
-
2\,\operatorname{tr}\!\left\{
\bigl(\Sigma^{1/2}\Sigma'\Sigma^{1/2}\bigr)^{1/2}
\right\}$,
which decomposes into contributions from the mean vectors and covariance matrices, with the latter inducing the Bures-Wasserstein metric $d \lo {\mathrm{BW}}$
on the space of SPD matrices.  The resulting metric space $(\mathcal S_{++}^2, d_{\mathrm{BW}})$ exhibits nonnegative Alexandrov curvature and, in general, does not admit an isometric embedding into a Hilbert space \citep{xu2025wasserstein}.

For the proposed DPE-based procedures, we use Gaussian kernels $\kappa(\cdot,\cdot)=\exp(-\gamma\|\cdot-\cdot\|^2)$ based on the squared DPE distances, with each bandwidth parameter $\gamma$ set to the inverse median of the corresponding pairwise squared distances, consistent with the heuristic adopted in \cite{gretton2007kernel}. For the conditional independence tests, we set $\epsilon_n$ in the Tikhonov regularization $(G_{\tilde Z}+n\epsn I)^{-1}$ to $0.005$.

\subsection{Unconditional Independence Testing}\label{sim: indep}

In this subsection, we evaluate the empirical performance of the DPE-based unconditional independence test and compare it with three existing methods: Profile Association (PA) \citep{zhou2025association}, Ball Covariance (Ball) \citep{wang_nonparametric_2024}, and generalized distance covariance (dCov) \citep{szekely2007measuring}.
The dCov, and Ball methods are implemented using the R packages \texttt{energy}, and \texttt{Ball}, respectively.
In each scenario, we vary the strength of dependence between $X$ and $Y$, as quantified by the parameter $0\le\rho\le1$. When $\rho=0$, $X$ and $Y$ are independent, so the empirical rejection rate estimates the type I error. As $\rho$ increases, we assess the power of the tests to detect the underlying dependence.

\begin{itemize}
    \item \textbf{Scenario 1 (Spherical Data)} We set the reference measures $\lambda_X$ and $\lambda_Y$ to be the normalized Euclidean surface measures on the corresponding unit spheres. Under this choice, the norm $\|\cdot\|_{\lambda_X}$ can be evaluated using the closed-form formula described in Section~S1.2 of the Supplementary Material.
For $i=1,\ldots,n$, we generate latent variables $\varepsilon_i=(\varepsilon_i^x,\varepsilon_i^y)\in\mathbb{R}^6$, where $\varepsilon_i^x,\varepsilon_i^y\in\mathbb{R}^3$. The entries of $\varepsilon_i$ are generated independently from the Gaussian mixture
\(
0.8\,N(1,1)+0.2\,N(-1,1).
\)
We define the predictor objects by
$X_i=\varepsilon_i^x/\|\varepsilon_i^x\|_2\in\mathbb{S}^2$
and construct the response objects by
$Y_i=\eta_i/\|\eta_i\|_2\in\mathbb{S}^2$,
where the intermediate vector $\eta_i\in\mathbb{R}^3$ is generated according to one of the following two settings:
\begin{itemize}
\item \textbf{Setting 1 (Multiplicative Dependence).} $\eta_{ij}=\bigl[\arccos(X_{ij})\bigr]^{\rho}\,\varepsilon^y_{i,j}, $ for $j=1,2,3$;
\item \textbf{Setting 2 (Additive Dependence).} $
\eta_{ij}=\rho |X_{ij}| + (1-\rho) \varepsilon^y_{i,j}, $ for $j=1,2,3$.
\end{itemize}

\medskip

 \item \textbf{Scenario 2 (SPD Matrices)}
Letting $\mathsf{Wishart}_p(\nu,\Sigma)$ denote the Wishart distribution with dimension $p$, degrees of freedom $\nu$, and   scale matrix $\Sigma $, we set the reference measures $\lambda_X$ and $\lambda_Y$ to be  $\mathsf{Wishart}_3(30,I_3)$.  We independently generate $X_i \simiid \mathsf{Wishart}_3(30,I_3)$ and $B_i \simiid \mathsf{Wishart}_3(30,I_3)$, and set {$F=30R$}, where $R$ is the fixed $3\times3$ equicorrelation matrix with diagonal entries one and off-diagonal entries $0.1$. The norm in \eqref{eq: Jx} is approximated by Monte Carlo integration using $n_{\mathrm{MC}}=500$ independent samples from the reference distribution $\mathsf{Wishart}_3(30,I_3)$. These Monte Carlo samples are generated independently of the observed data. We consider the following dependence structures:
\begin{itemize}
\item \textbf{Setting 1 (Multiplicative Dependence).}
\quad $
Y_i=(10\,d_R(X_i,F))^{\rho}\,F B_i F.
$

\item \textbf{Setting 2 (Additive Dependence).} \quad
$
Y_i=0.5\,\rho\, d_R(X_i,F)\,F+(1-\rho)B_i.
$
\end{itemize}

\medskip

 \item \textbf{Scenario 3 (Bivariate Gaussian Distributions)}
In this case,  $(X,Y)$ is a pair of dependent random bivariate Gaussian distributions. Specifically, $X$ is the random distribution $N(\mu \hi X , \Sigma \hi X)$, where $\mu \hi X  \sim N(0, 5^2 I \lo 2)$ and
$\Sigma \hi X=\bigl(\begin{smallmatrix}1&U\\U&1\end{smallmatrix}\bigr)$,
where $U \sim \mbox{Unif}(0, 0.5)$. The random distribution $Y$ is the bivariate Gaussian $N(\mu \hi Y, \Sigma \hi Y)$, where the random quantities $\mu \hi Y$ and/or $\Sigma \hi Y$ are statistically dependent on $X$. Let $(X \lo 1, Y \lo 1), \ldots, (X \lo n, Y \lo n)$ be independent copies of $(X,Y)$. We do not observe this sample. Instead, we observe 2-dimensional random vectors $X \lo {i1}, \ldots, X \lo {iq}$ and $Y \lo {i1}, \ldots, Y \lo {iq}$ such that
   $( X \lo {i1}, \ldots, X \lo {iq} | X \lo i) \sim X \lo i$ and  $( Y \lo {i1}, \ldots, Y \lo {iq} | Y \lo i) \sim Y \lo i$.
We take $q=\min(n,100)$ and use the sample means and covariance matrices to evaluate the Wasserstein distances and construct the DPE kernels. We approximate the reference-measure integrals using 500 Gaussian distributions $N(\mu_r,\Sigma_r)$ generated independently of the observed data, where $\mu_r\sim N(0,I_2)$ and $\Sigma_r\sim\mathsf{Wishart}_2(2,I_2/5)$.
Following  \cite{wang_nonparametric_2024}, we consider   two dependence structures of $Y$ on $X$, one driven by mean and one driven by variance. The strength of dependence is controlled by the parameter $\rho \in [0,1]$, with larger $\rho$ representing strong dependence, smaller $\rho$ representing weak dependence, and $\rho=0$ representing independence.

\medskip

\def\trans{^\top}

\begin{itemize}
    \item \textbf{Setting 1 (Mean-Driven Dependence).} We set $\mu \hi Y = \rho (V \lo 1, V \lo 2)\trans + (1-\rho) \epsilon$, where
 $ V \lo 1=  W_2 \left(  N(\mu^X ,\Sigma^X ),  N(0,I_2)\right)$,
 $ V \lo 2 =  W_2 \left( N(\mu^X,\Sigma^X), N((-1, 1) \trans ,I_2)\right)$,  and
  $\epsilon \sim N(0,5^2 I_2)$.
The matrix $\Sigma \hi Y$ is set to the nonrandom value
$\bigl(\begin{smallmatrix}1&-0.5\\-0.5&1\end{smallmatrix}\bigr)$.
Thus, the random distribution $Y$ depends on $X$ only through its random mean $\mu \hi Y$.
\vspace{.08in}
    \item  \textbf{Setting 2 (Variance-Driven Dependence).}
Let \(d_{X1,i}=W_2\!\left(\mathcal N(\mu_i^X,\Sigma_i^X),\mathcal N(0,I_2)\right)\), set {\(A=\bigl(\begin{smallmatrix}0.2&0.2^2\\0.2^2&0.2\end{smallmatrix}\bigr)\)}, and draw {\(B_i\simiid\mathsf{Wishart}_2(2,I_2/5)\) independently of \(X_i\)}. We set \(\mu_i^Y\equiv0\) and define {\(\Sigma_i^Y=0.1\rho d_{X1,i}A+(1-\rho)B_i\)}.
\end{itemize}
\end{itemize}

\vspace{.04in}
For each scenario, we report empirical power curves based on 500 Monte Carlo replications, with the significance level fixed at $0.05$. We examine rejection probabilities over $\rho\in\{0,0.1,\ldots,1\}$ at sample sizes $n\in\{50,100,200\}$ for Scenario~1 and $n\in\{50,100\}$ for Scenarios~2 and~3.

\begin{figure}[htpb]
    \centering
    \includegraphics[width=0.8\linewidth]{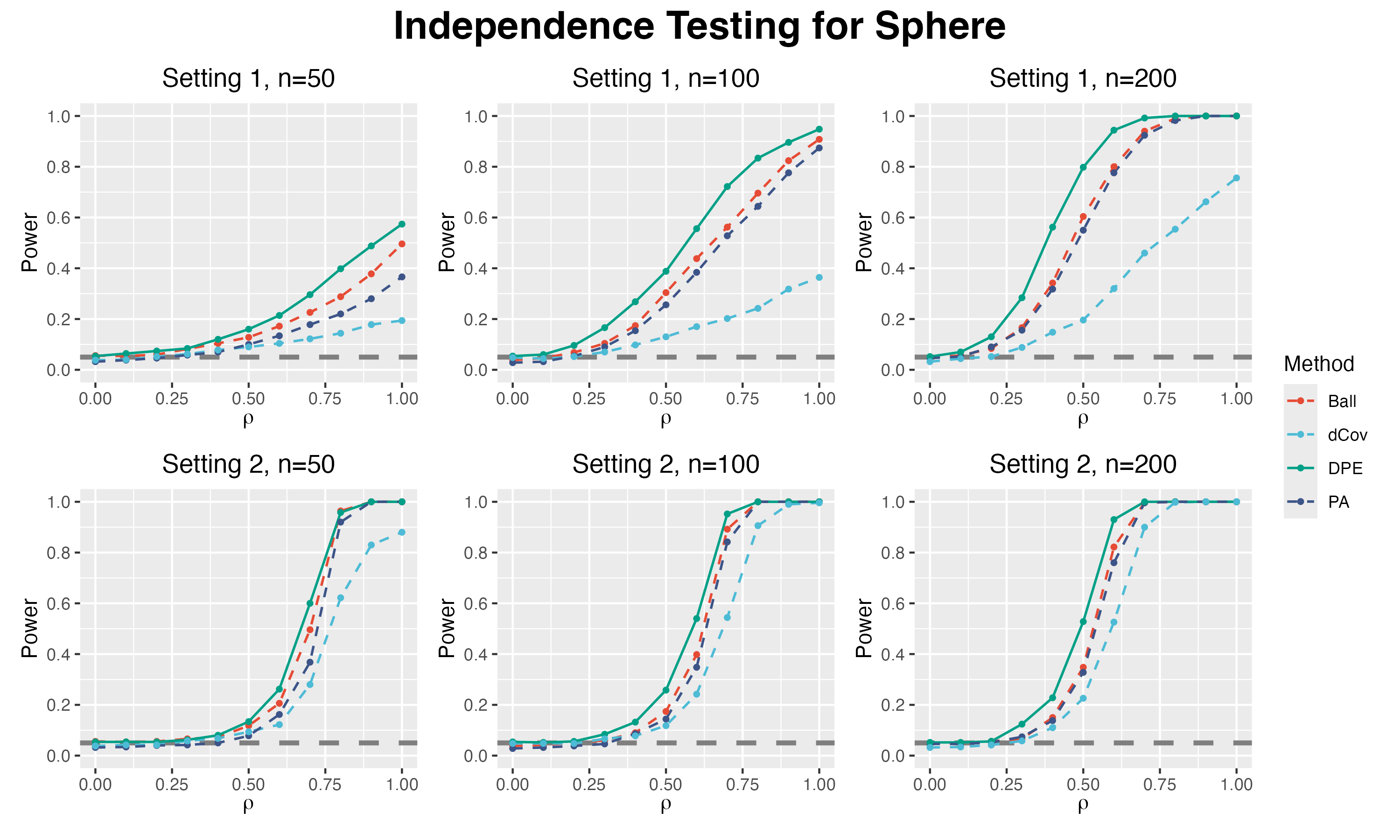}
    \caption{Empirical power for testing independence on $\mathbb{S}^2$ (Scenario 1) as a function of the dependence strength $\rho$ at the significance level $0.05$. Results are based on 500 Monte Carlo replications with sample sizes $n \in \{50, 100, 200\}$. The grey horizontal dashed line denotes the nominal type I error rate.}
    \label{fig:pw S2}
\end{figure}

\begin{figure}[htpb]
    \centering
    \begin{subfigure}[b]{0.49\textwidth}
        \centering
        \includegraphics[width=\linewidth]{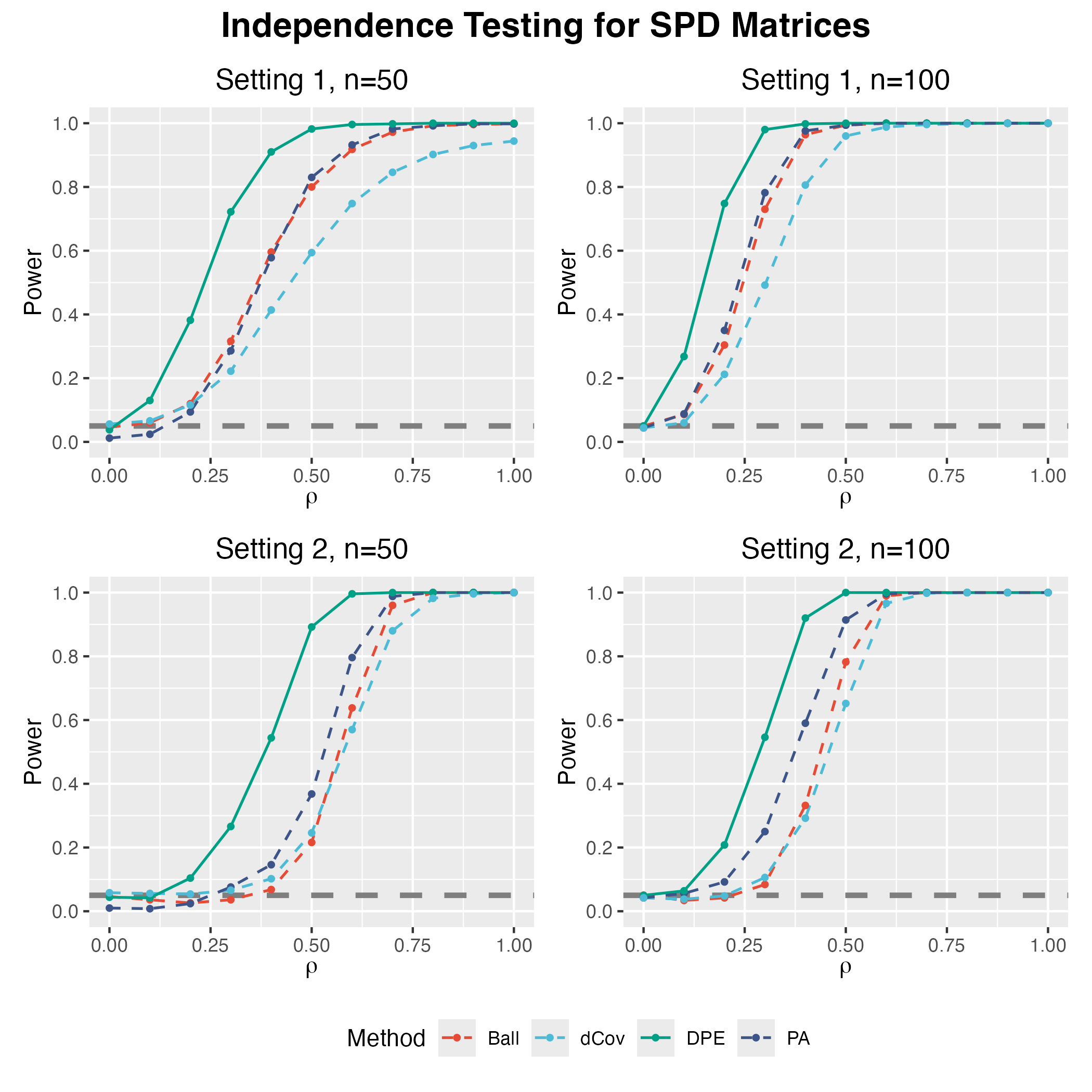}
        \caption{SPD matrices (Scenario 2)}
        \label{fig:pw SPD}
    \end{subfigure}
    \hfill
    \begin{subfigure}[b]{0.49\textwidth}
        \centering
        \includegraphics[width=\linewidth]{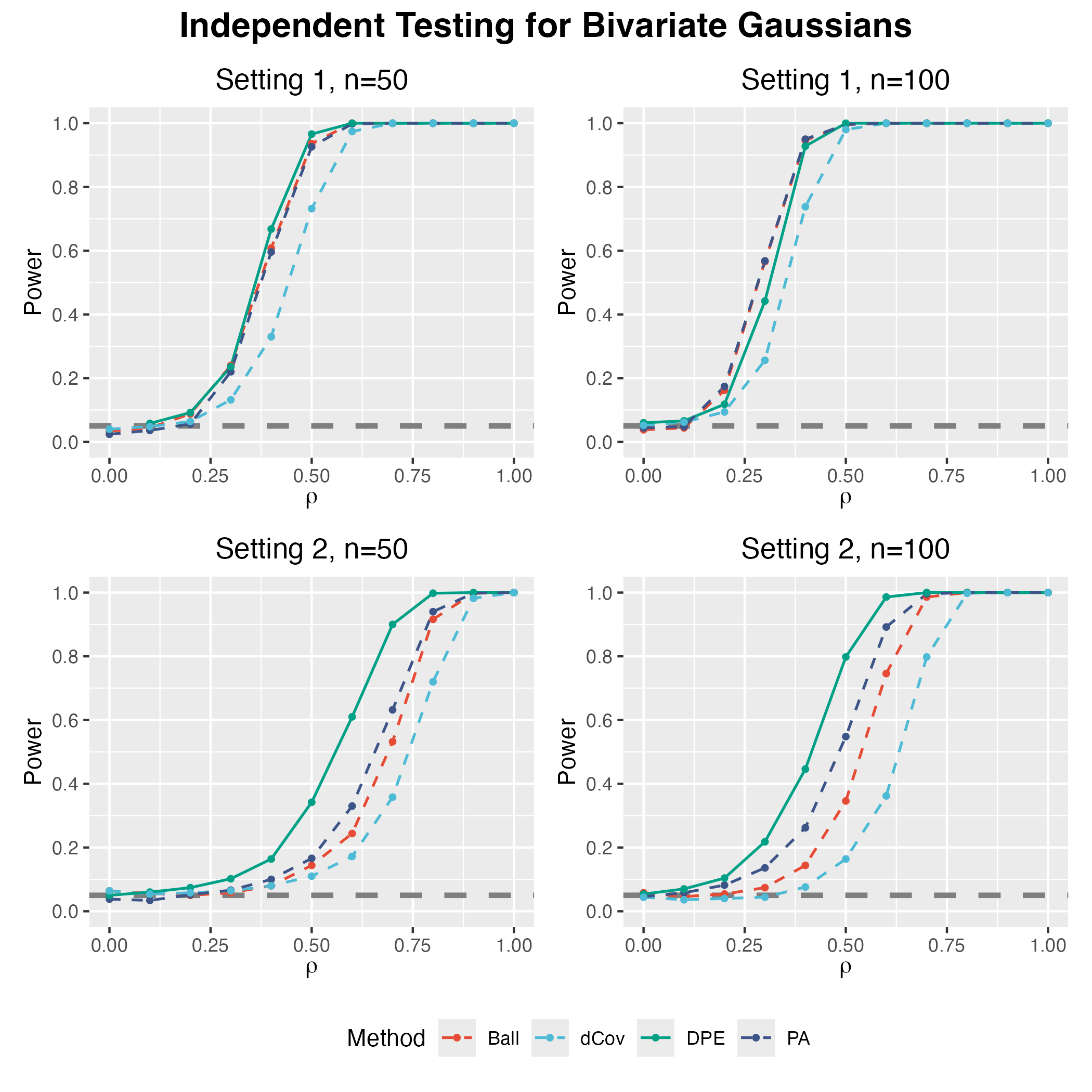}
        \caption{Bivariate Gaussians (Scenario 3)}
        \label{fig:pw G2}
    \end{subfigure}

    \caption{Empirical power for testing independence on SPD matrices (Scenario 2) and bivariate Gaussians (Scenario 3) as a function of the dependence strength $\rho$ at a significance level $0.05$. Results are based on 500 Monte Carlo replications with sample sizes $n \in \{50, 100\}$. The grey horizontal dashed line denotes the nominal type I error rate.}
    \label{fig:power_combined}
\end{figure}

As shown in Figure~\ref{fig:pw S2}, all methods maintain type I error rates close to the nominal level of $0.05$. Consistent with our asymptotic theory, the empirical power of the DPE-based test increases with $n$ at each fixed value of $\rho$ in each setting. These results indicate that our DPE framework effectively captures   non-Euclidean dependence structures whereas traditional approaches, such as dCov, often exhibit a loss of power. Specifically, among the four tests, the proposed DPE method consistently achieves the highest power across both settings in Scenario 1. As shown in Figure~\ref{fig:pw SPD}, Scenario~2 exhibits a similar pattern, with DPE again achieving the highest power. For Scenario 3 in Figure~\ref{fig:pw G2}, in Setting~1, where dependence is driven by the means of the distributions, Ball, PA, and DPE exhibit comparable power across the full range of $\rho$. In Setting~2, where the random objects are dependent through their variances, the proposed DPE test significantly outperforms the competing methods,   suggesting that the DPE is more adept at capturing second-moment dependencies in the Wasserstein space.

\subsection{Conditional Independence Testing}
We now turn to conditional independence tests. To our knowledge, there are currently no methods specifically designed to test the conditional independence between $X$ and $Y$ given a general non-Euclidean variable $Z$. We evaluate the performance of the proposed method across the three scenarios described previously. The models for $(X, Y, Z)$ are constructed on the same non-Hilbertian geometries as those used in the unconditional case. However, the sampling strategies are modified to distinguish between conditional and unconditional associations rigorously. In all scenarios, we ensure that $X$ and $Y$ exhibit unconditional dependence to test the statistic's ability to remove the influence of $Z$, and we use a positive $\rho$ to control the strength of the unconditional dependence between $(X,Y)$. In each scenario, Setting~1 is constructed so that the null hypothesis
$X \indep Y \mid Z$ holds, allowing us to evaluate empirical type I
error control. Setting~2 is  constructed so that
$X \indep Y \mid Z$ fails, allowing us to assess the power of the test. The tuning parameter $\epsn$ in the Tikhonov regularization $(G_{\tilde Z}+n\epsn I)^{-1}$ is fixed at $0.005$, which provides a stable numerical approximation of the inverse while being consistent with our theory.

\begin{itemize}
    \item \textbf{Scenario 1 (Spherical Data)} We generate $n$ independent random vectors \(\varepsilon_i=(\varepsilon_i^x,\varepsilon_i^y)\in\mathbb R^6\),   \(i=1,\dots,n\), where the six entries in $\varepsilon_i $  are  \emph{i.i.d.} and follow the Gaussian mixture \(0.8   N(1,1) + 0.2  N(-1,1)\). The conditioning variable $Z_i$ follows a 3-dimensional standard logistic normal distribution. The observed random objects are
$X_i=\frac{\eta_i^x}{\|\eta_i^x\|_2}$ and
$Y_i=\frac{\eta_i^y}{\|\eta_i^y\|_2}$,
taking values in $\mathbb{S}^2$, where $\eta_i^x,\eta_i^y\in\mathbb{R}^3$ are generated according to the following settings:
\begin{itemize}
    \item \textbf{Setting 1.}
Let
$\eta_{ij}^x=\bigl[\arccos(Z_{ij})\bigr]^{\rho}\varepsilon_{ij}^x$ and
$\eta_{ij}^y=\bigl[\arccos(Z_{ij})\bigr]^{\rho}\varepsilon_{ij}^y$, for $j=1,2,3$. By construction, $X_i \indep Y_i \mid Z_i$.
\item \textbf{Setting 2.}
Let
$\eta_{ij}^y=\arccos(Z_{ij})\,\varepsilon_{ij}^y$ and
$\eta_{ij}^x=\arccos(Z_{ij})
\bigl[\rho\,\varepsilon_{ij}^y+(1-\rho)\,\varepsilon_{ij}^x\bigr]$, for $j=1,2,3$. Under this construction, $X_i  \indep Y_i \mid Z_i$ fails whenever $\rho>0$.
\end{itemize}

\medskip
    \item \textbf{Scenario 2 (SPD Matrices)}  \quad In this scenario, $X$ and $Y$ are SPD matrices and $Z$ is a random object in $\mathbb{S} \hi 2$. We use the same Monte Carlo approximation procedure  for the $L \lo 2 ( \Omega \lo X, \lambda \lo X)$ and $L \lo 2 (\Omega \lo Y, \lambda \lo Y)$ norms as in Scenario 2 of Section \ref{sim: indep}, and the reference measure $\lambda_Z$ is defined as the uniform distribution over the sphere $\mathbb{S}^2$. We sample $A \lo 1, \ldots, A \lo n, B \lo 1, \ldots, B \lo n$ independently from  $\mathsf{Wishart}_3(30,I_3)$ and generate $Z_i = |\varepsilon^z_i| / \|\varepsilon^z_i\|_2 \in \mathbb{S}^2$, where $\varepsilon \hi z \lo 1, \ldots, \varepsilon^z_n$ are independently sampled from $N (0,I_3)$. Let $F_i=30\diag(Z_i)\diag(Z_i)$.
\begin{itemize}
    \item \textbf{Setting 1.}   $X_i=\rho F_i+(1-\rho)A_i$, and $
Y_i=\rho F_i+(1-\rho)B_i$. While $X_i$ and $Y_i$ are unconditionally dependent due to the common term $F_i$, they are independent given $Z_i$.
\item \textbf{Setting 2.} $X_i=A_i$, and $Y_i=5\rho\,d_R(A_i,30I_3)\,F_i+(1-\rho)B_i$. This creates conditional dependence where the Riemannian geometry of $X_i$ influences the scale of $Y_i$.
\end{itemize}
\medskip

    \item \textbf{Scenario 3 (Bivariate Gaussians)} \quad Here, $X, Y$ are bivariate Gaussian distributions, and $Z$ is a random matrix. To calculate \eqref{eq: Jx} for $X, Y$, we use $n_{\mathrm{MC}} = 500$ samples from $N(\mu_r, \Sigma_r)$ with $\mu_r \sim N(0, I_2)$ and $\Sigma_r \sim \mathsf{Wishart}_2(20,\;I_2)$ for Monte Carlo approximation. Similarly, we use $n_{\mathrm{MC}} = 500$ samples from $\mathsf{Wishart}_2(30,\;I_2)$ to approximate the $\lz$-integral. For $i=1,\ldots,n$, we draw $Z_i\simiid\mathsf{Wishart}_2(10,I_2)$ and define $X_i=N(0,\Sigma_i^X)$ and $Y_i=N(0,\Sigma_i^Y)$. The covariance matrices $\Sigma_i^X$ and $\Sigma_i^Y$ are specified by the following settings.
\begin{itemize}
    \item \textbf{Setting 1.} Set \(\Sigma_i^X=Z_i+\tilde\Sigma_i\), where
\(\tilde\Sigma_i\simiid\mathsf{Wishart}_2(10,I_2)\) independently of \(Z_i\). Let
\(\eta=(-1,1)^\top\),
\(\Lambda_1=\bigl(\begin{smallmatrix}1&-1\\-1&1\end{smallmatrix}\bigr)\), and
\(\Lambda_2=\bigl(\begin{smallmatrix}1&1\\1&1\end{smallmatrix}\bigr)\).
Independently draw \(B_i\simiid\mathsf{Wishart}_2(10,I_2)\) and define
\(\Sigma_i^Y=\rho\,\mathrm{diag}(U_{1i},U_{2i})+(1-\rho)B_i\), where
\(U_{1i}\) and \(U_{2i}\) are the \(W_2\) distances from \(N(0,Z_i)\) to
\(N(0,\Lambda_1)\) and \(N(\eta,\Lambda_2)\), respectively.
    \item \textbf{Setting 2.} Draw
\(\Sigma_i^X\simiid\mathsf{Wishart}_2(10,I_2)\) independently of \(Z_i\), and define
\(\Sigma_i^Y=\rho\,\mathrm{diag}(V_{1i},V_{2i})+(1-\rho)Z_i\), where
\(V_{1i}\) and \(V_{2i}\) are the \(W_2\) distances from \(N(0,\Sigma_i^X)\) to
\(N(0,\Lambda_1)\) and \(N(\eta,\Lambda_2)\), respectively.
\end{itemize}
\end{itemize}

\begin{figure}[htp]
    \centering
    \includegraphics[width=0.8\linewidth]{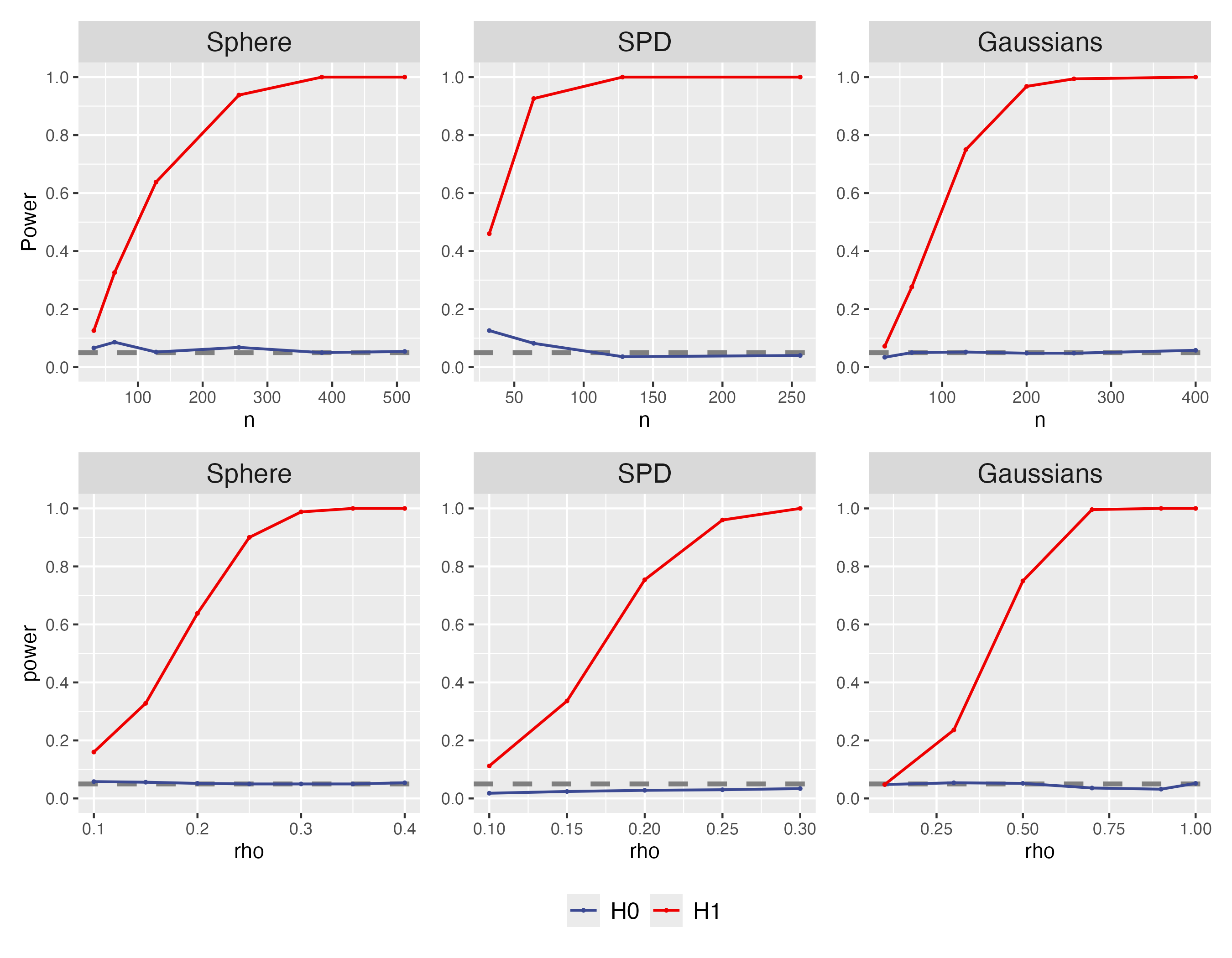}
    \caption{Empirical rejection rates for the proposed DPE-based conditional independence test. Blue and red curves denote performance under the null hypothesis ($H_0$) and the alternative ($H_1$), respectively. The black horizontal dashed line represents the nominal level $0.05$. In the upper panels,  the dependence strength $\rho$ between $X$ and $Y$ is fixed as the sample size $n$ grows. In the lower panels, $\rho$ increases while the sample size is fixed at $n=128$. }
    \label{fig:pw CI}
\end{figure}

Figure~\ref{fig:pw CI} shows that the proposed conditional independence test approaches the nominal type I error level as the sample size grows while achieving increasing power under the alternatives. The upper panels display the rejection rates as a function of sample size $n$ with the dependence strength fixed at $\rho = 0.5, 0.4,$ and $0.2$ for the Gaussian, SPD, and Spherical scenarios, respectively. The lower panels present the power and size as functions of $\rho$ with the sample size fixed at $n=128$.

Under the null hypothesis ($H_0$, blue curves), the empirical size is generally near the nominal level $0.05$. The SPD scenario exhibits finite-sample size inflation at the two smallest sample sizes, but the test achieves stable size control once $n$ exceeds 100, consistent with the asymptotic validity established in Theorem~\ref{thm: null dist CI}. These results confirm that our DPE framework successfully identifies conditional independence even when $X$ and $Y$ exhibit strong unconditional dependence, demonstrating its ability to effectively adjust for the confounding influence of the object-valued variable $Z$.

Under the alternative hypothesis ($H_1$, red curves), the empirical power increases and converges rapidly to 1 as a function of $n$ and $\rho$. This high sensitivity to conditional dependence is consistently observed across all scenarios, validating the effectiveness of our DPE framework in non-Euclidean settings where the conditioning variable has a complex manifold structure.

We also investigate performance under local alternatives in which the cross-covariance operator $\sigt$ or the conditional cross-covariance operator $\Sigxyz$ shrinks to zero, i.e. $\|\sigt\|_{\HS}\to 0$ and $\|\Sigxyz\|_{\HS}\to0$. The empirical results, reported in Section~S3.1 of the Supplementary Material, are consistent with the theoretical characterizations in Theorems~\ref{thm: power} and~\ref{thm: power CI}, and demonstrate that the DPE-based test remains powerful for detecting vanishing dependencies in object-valued data.

\section{Real Applications}\label{sec: rda}
To show the practical utility of our DPE-based testing framework, we apply our methods to two real-world problems involving complex random objects.

\subsection{Microbiome Data: Marginal Independence Testing}\label{microbiome rda}
The human gut microbiome is vital in nutrient digestion and metabolic regulation \citep{COMBO2011Gary}, and extensive research has reported a significant association between overall gut microbial community structure and BMI \citep{tang2017,tan2025high}. In the COMBO study \citep{COMBO2011Gary}, gut microbiome compositions are measured as relative abundances of operational taxonomic units (OTUs). These observations reside on the compositional simplex $\mathcal S_{\ge 0}^{p-1}=\{(x_1,\dots,x_p): \sum_{j=1}^p x_j=1,\ 0\le x_j\le 1\}$.
After standard data filtering and quality control, we retained $n=91$ samples for analysis. To evaluate the dependence between microbiome structure and BMI, we consider four metrics that capture distinct structural features of taxonomic data: UniFrac, generalized UniFrac (GUniFrac), weighted UniFrac (WUniFrac), and Aitchison distances   \citep{li_microbiome_2015}. These metrics differ significantly in their geometric properties: the Aitchison distance isometrically embeds compositions into a log-ratio Euclidean space, thereby permitting the application of standard Euclidean methodologies, and in contrast, UniFrac-based distances incorporate phylogenetic tree structures to aggregate mass differences between samples and thus are not isometrically embeddable into a Hilbert space because they fail to satisfy the negative-type property \citep{zhu2025mathematical}.

\begin{table}[ht]
\centering
\caption{Comparison of $p$-values for marginal independence testing between BMI and gut microbiome community structure on the COMBO dataset. The results compare the proposed DPE framework against benchmarks across four distance metrics. The  “Embeddable” column indicates whether the metric allows the compositional simplex to be isometrically embedded into a Hilbert space.}
\begin{tabular}{cc|cccc}
\hline
 Metric&  Embeddable & Ball & dCov & PA & DPE \\
\hline
Aitchison& $\checkmark$ & 0.0400 & 0.0050 & 0.0398 & 0.0052 \\
UniFrac & $\times$  & 0.0067 & 0.0050 & 0.0149 & 0.0120 \\
GUniFrac& $\times$  & 0.0600 & 0.0250 & 0.0746 & 0.0334 \\
WUniFrac & $\times$ & 0.0400 & 0.0950 & 0.0448 & 0.0324 \\
\hline
\end{tabular}
\label{tab:microbiome}
\end{table}

To compute \eqref{eq: Jx} on the simplex $\c S_{\ge 0}^{p-1}$, we employ the empirical measure $\lx=n^{-1}\sum^n_{i=1}\delta_{X_i}$, where $\delta_a$ denotes the Dirac measure at $a$. Table~\ref{tab:microbiome} reports the resulting $p$-values for the DPE-based test and its competitors across the four metrics. For the Aitchison distance, which isometrically embeds into a Hilbert space, all methods yield statistically significant results at the 0.05 level, confirming the presence of a strong marginal association. However, for the UniFrac-based metrics that incorporate phylogenetic tree structures, the performance of existing methods deteriorates. Only the proposed DPE-based test consistently detects a significant dependence across all UniFrac-based distances. Unlike traditional distance-based kernels that implicitly rely on the negative-type property, the DPE provides a rigorous and unified approach that remains valid across both Aitchison and UniFrac-based geometries.

\subsection{Human Mortality Data: Conditional Independence Testing}\label{mortality rda}
Historically, the biological and systemic risks associated with reproductive history have been pivotal determinants of female longevity. We study the association between the distribution of female age-at-death and the distribution of maternal age at childbirth using the proposed distributional conditional independence framework. To isolate reproductive-specific health effects from shared environmental or socioeconomic stressors that affect a population broadly, we test whether the female mortality distribution is conditionally independent of the maternal fertility distribution, given the male age-at-death distribution. In this context, the male age-at-death distribution serves as a robust proxy for the common mortality environment and cohort-specific external risks.

We utilize the UN World Population Prospects 2019 database (\url{https://population.un.org}), which provides internationally comparable estimates of fertility and mortality for $n=201$ countries. For each country and five-year calendar period, we construct age-specific distributions of deaths (ages 0--100) and births (maternal ages 15--50) using smoothed life tables. From a statistical perspective, this problem is uniquely non-trivial: all random elements involved are probability distributions residing in the non-Euclidean  Wasserstein space. This structure renders traditional conditional independence tests for vector-valued data inapplicable and necessitates the flexible embedding provided by the DPE.

\begin{figure}[ht]
    \centering
\includegraphics[width=0.66\linewidth]{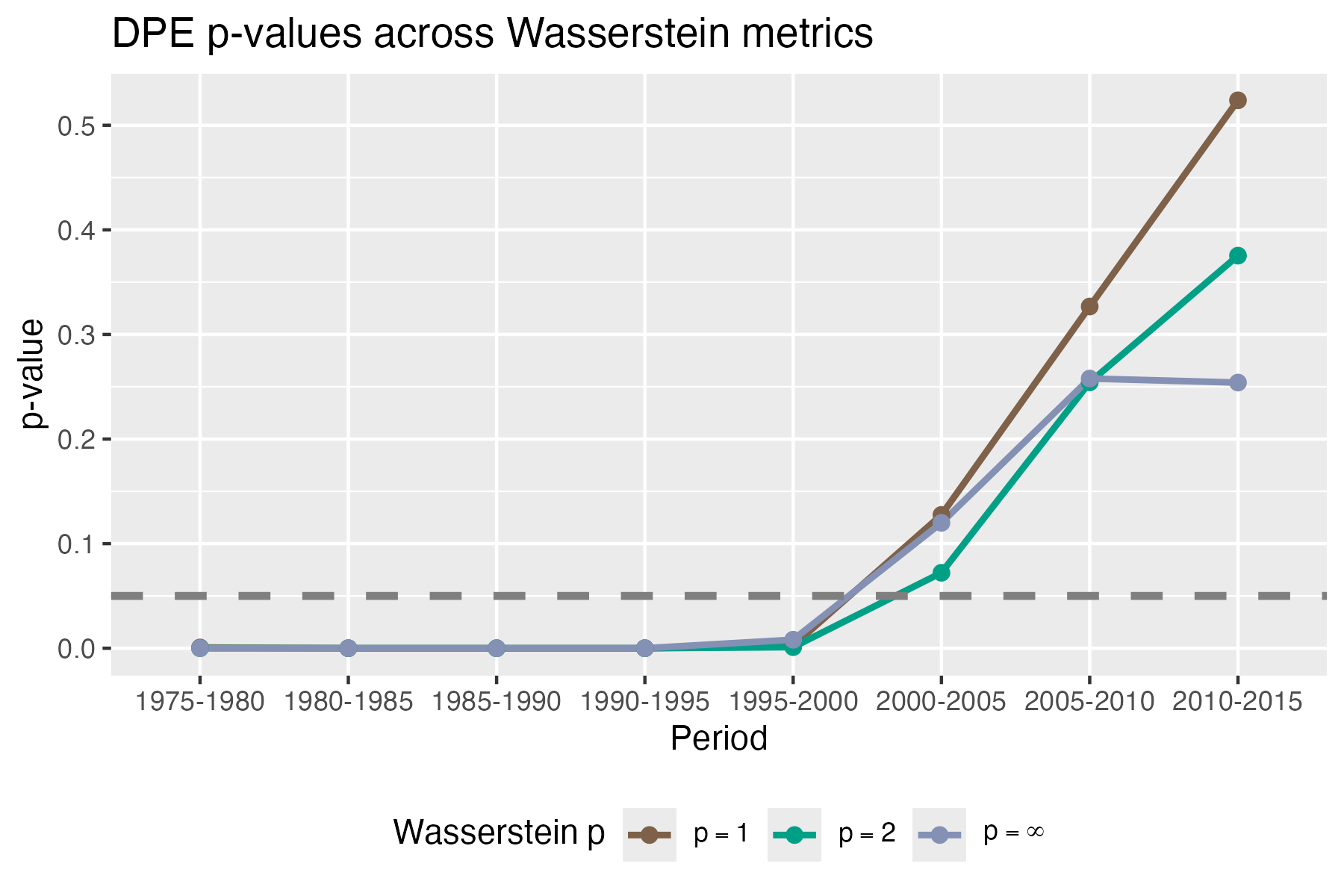}
    \caption{Empirical $p$-values for conditional independence testing between female mortality and fertility distributions given male mortality distribution. The curves show the proposed DPE-based test evaluated across five-year intervals under different Wasserstein$-p$ metrics. The dashed horizontal line denotes the significance level $0.05$.}
    \label{fig:mortality}
\end{figure}

We apply the proposed DPE-based conditional independence test to these distributional observations under each Wasserstein metric. To examine the sensitivity of our results to the choice of metric, we treat these distributions as probability measures in the Wasserstein-$p$ space equipped with the Wasserstein-$p$ distance with $p \in \{1, 2, \infty\}$. The required $L \lo 2(\Omega_X,\lambda_X)$ distances are approximated via Monte Carlo integration, using standard Brownian motion trajectories scaled by a factor of 100. Figure~\ref{fig:mortality} reports the resulting $p$-values for the conditional independence tests across five-year periods. We have the following interesting findings. During 1980--1995, the $p$-values are uniformly close to zero across all choices of $p$, indicating a significant residual dependence between female mortality and fertility timing, even after conditioning on the shared mortality environment. However, the $p$-values increase steadily from the late 1990s and exceed the 5\% significance level after 2000, suggesting a progressive temporal decoupling of the conditional fertility-mortality association. This temporal shift is consistent with global declines in pregnancy-related mortality and the widespread expansion of obstetric care and reproductive health services since the 1990s \citep{kassebaum2014global,WHO2019MaternalMortality}. These results show the ability of our DPE framework to detect meaningful conditional dependence structures in distributional data while remaining robust to the choice of the Wasserstein metric.

\section{Conclusion}\label{conclusion}

This paper introduces the Distance Profile Embedding (DPE), a novel framework for testing independence and conditional independence between random objects in a general metric space. To our knowledge, DPE is the first general framework for testing independence and conditional independence between random objects in arbitrary metric spaces without imposing additional structural assumptions on the metric space, such as negative type or isometric embeddability. The key idea is to embed the random objects into a Hilbert space of distance profiles, though not necessarily isometrically. Because this embedding is measurable and injective, it preserves both independence and conditional independence.

\section*{Supplementary Material}
\noindent
The supplement provides the full proofs, lemmas, and additional simulation results.

\bibliographystyle{agsm}
\bibliography{ref}
\end{document}